# A Hybrid Dependency-Aware Framework for Task Decomposition and Dynamic Agent Generation in Oracle-to-PostgreSQL Migration

Oleg Grynets
EPAM Systems
McLean, Virginia, USA
oleg_grynets@epam.com

Oleg Kaskun
EPAM Systems
Lviv, Ukraine
oleh_kaskun@epam.com

Alona Seletska
EPAM Systems
Kyiv, Ukraine
alona_seletska@epam.com

Daryna Tukalo
EPAM Systems
Kyiv, Ukraine
daryna_tukalo@epam.com

Vasyl Lyashkevych
EPAM Systems
Lviv, Ukraine
vasyl_lyashkevych@epam.com

***Abstract*—Large language model (LLM)-based database migration is usually treated as a direct code transformation problem, although enterprise Oracle estates contain heterogeneous SQL and PL/SQL artifacts whose dependencies, execution order, migration complexity, and validation requirements differ substantially. This paper proposes a hybrid dependency-aware framework that first identifies migration tasks, then builds a cross-file dependency graph, condenses cyclic dependencies, and uses the resulting task specifications to drive runtime generation of specialized migration agents. The deterministic path uses ANTLR-based parsing and typed dependency extraction, while an LLM fallback is activated only for units that cannot be parsed reliably. On a real corpus of 116 Oracle files, the deterministic pipeline produced 1,037 units with zero coverage gaps and 1,271 AST-derived dependencies. The selective fallback processed 165 parse-error units, recovered 496 additional validated dependencies, reduced unresolved-dependency units from 165 to zero, and increased resolved internal edges from 446 to 527. The resulting graph contained four dependency-respecting phases; cycle handling was separately validated on a controlled fixture by Tarjan strongly connected component condensation. A complementary end-to-end migration experiment on 1,006 PL/SQL files produced 623 regenerated scripts (about 62%); 380 of those 623 scripts (about 61%) executed successfully in PostgreSQL 16. Tables reached approximately 85% regeneration success, while no query regenerations succeeded under the evaluated specification-mediated baseline, and procedural objects remained strongly dependent on schema context. These results motivate dependency-aware context delivery and task-specific verification rather than a single uniform migration agent. The paper formalizes task-to-agent mapping, introduces a monitoring and diagnostic layer for execution feedback, and defines a controlled comparison of monolithic, static-decomposition, dependency-aware, and dynamically orchestrated migration strategies. The reported migration experiment is used as an empirical baseline and motivation; it does not by itself prove that dynamic-agent routing is superior, which remains the target of the comparative experiment.***

## I. Introduction

### A. Background and Motivation

LLMs are now used for code generation, program repair, software modernization, and repository-level engineering, but reliability remains difficult when tasks require long contexts, tool use, multi-step decisions, and cross-artifact reasoning [1], [2]. Long-context studies show that simply enlarging the context window does not guarantee robust use of relevant information, while prompt-compression methods reduce cost but do not by themselves recover the structural dependencies of a software system [3]–[5]. These limitations are particularly important in Oracle-to-PostgreSQL migration, where a single migration unit can depend on tables, views, packages, sequences, procedures, triggers, types, cursors, and implicit execution semantics distributed across many files.

Previous work on token optimization for Oracle-to-PostgreSQL migration demonstrated that context reduction must be treated as a multi-objective problem rather than as simple prompt shortening [6]. Specification-based reengineering further showed that controlled software transformation benefits from explicit intermediate representations and traceability rather than direct Code2Code generation [7]. Related specification-driven monitoring studies also support the idea that LLM-generated artifacts should be governed by explicit, regenerable specifications and monitored under system evolution [8]–[10]. The next step is therefore structural: before deciding how much context to give an agent, the system must identify what the migration tasks are, which tasks depend on which others, and which execution structure is appropriate for each task.

Agent frameworks and programmable LM pipelines such as AutoGen, DSPy, and DSPy Assertions support multi-step execution and constrained inference workflows [11]–[13], while recent work on runtime-structured decomposition, dynamic task decomposition, agent evaluation, software-engineering benchmarks, and multi-agent orchestration highlights the importance of workflow structure, reliability, and execution control [14]–[20]. However, most agentic decomposition research starts from an already defined task or query graph. It does not address how a graph should be derived from heterogeneous legacy database source artifacts, nor how a derived dependency graph can become an operational specification for creating specialized migration agents at runtime.

This paper addresses that gap with a deterministic-first hybrid framework. Oracle artifacts are decomposed into typed units, dependencies are extracted across files, unresolved parse-error units are selectively analyzed by an LLM, strongly connected components are condensed into cycle work items, and the resulting DAG exposes a migration order and parallel phases. Each node is then converted into an agent-ready task specification whose type, complexity, dependency closure, risk, and acceptance criteria determine the runtime agent configuration.

A complementary migration-generation study provides an important execution-level signal for this architecture. A specification-first pipeline was applied to 1,006 PL/SQL files; 623 files were regenerated and 380 of the regenerated scripts executed successfully in a PostgreSQL 16 Docker environment. The large difference between structural objects and procedural or cross-object artifacts indicates that migration success is not uniform across task classes. In particular, the supplied experiment reports approximately 85% success for tables, no successful query regenerations under the evaluated specification-mediated baseline, and recurrent failures of functions, procedures, and indexes when the target schema context was incomplete. These observations do not constitute an ablation of dependency-aware routing, but they define a concrete failure surface that the proposed dependency graph and dynamic-agent layer are designed to address. Accordingly, the central research question of this study is whether a hybrid dependency-aware task decomposition framework, combined with task-conditioned dynamic agent generation, can provide a more reliable, efficient, and recoverable basis for Oracle-to-PostgreSQL migration than monolithic or statically decomposed LLM-based approaches.

The contributions are fourfold. First, we define a hybrid decomposition model that preserves deterministic grammar-derived structure and uses an LLM only as a validated fallback. Second, we transform decomposition output into a cross-file dependency graph with explicit external references, cycle condensation, order, and parallel phases. Third, we introduce a task-to-agent mapping that supports dynamic creation of single agents, agent-validator pairs, or temporary multi-agent teams. Fourth, we report measured decomposition results on a 116-file Oracle corpus and complementary execution-level results from 1,006 PL/SQL files, while defining a controlled C0–C3 experiment for isolating the effects of dependency-aware context and dynamic agent generation.

## II. Related Work

### A. LLM-Assisted Software and Database Migration

LLM-assisted software migration has evolved from direct code translation toward structured transformation pipelines that combine source analysis, intermediate representations, retrieval, dependency handling, verification, and target-specific validation. For Oracle-to-PostgreSQL migration, the principal challenge is not only dialect translation but also preservation of cross-object dependencies, execution order, schema assumptions, and traceability across heterogeneous SQL and PL/SQL artifacts [6], [7], [21].

These studies suggest a useful architectural separation for database migration: probabilistic components can propose transformations, whereas deterministic components should preserve source structure, resolve dependencies, enforce validation gates, and record execution evidence. Accordingly, the proposed framework confines LLM use to selective dependency recovery and migration generation, while source coverage checking, graph construction, SCC condensation, and target-side execution validation remain explicitly controlled [6], [7], [21].

### B. Task Decomposition and Agent Orchestration

AutoGen supports customizable conversable agents [11], DSPy treats LM calls as composable modules [12], and DSPy Assertions adds computational constraints and inference-time refinement [13]. LLMCompiler targets parallel function calling [22], while Reflexion uses feedback and episodic memory for iterative improvement [23]. AgentBench provides a multi-environment evaluation of agent behavior [16]. These works establish important orchestration mechanisms, but the workflow structure is generally supplied by the developer or by a planner operating over an already abstract task.

Runtime-Structured Task Decomposition (RSTD) explicitly separates deterministic orchestration from narrow LLM judgment calls. In Kubernetes root-cause analysis, RSTD reduced simulated retry cost from 1,632 ± 145 tokens for static decomposition to 436 ± 132 tokens, illustrating that decomposition without runtime isolation can increase rather than decrease retry cost [14]. TDAG goes further by combining dynamic task decomposition with on-the-fly agent generation [15]. Modular task decomposition and dynamic collaboration introduces solvability, non-redundancy, scheduling, and global consistency constraints [19]. VMAO represents subtasks as a DAG and couples execution with verification-driven replanning [20]. AdaptOrch selects execution topology based on a task dependency graph [24], and recent surveys formalize such workflows as agentic computation graphs whose structure can be static or dynamically modified [25].

### C. Specification-Driven and Dependency-Aware Execution

Recent spec-driven development treats the specification as an operational source of truth rather than as passive documentation [26], extending earlier specification-driven approaches that integrated executable specifications, contracts, and agile development processes [27]. This view aligns naturally with migration tasks because each decomposed unit can carry an explicit objective, dependencies, constraints, acceptance criteria, and routing hints. In adjacent domains, task decomposition plus verification has improved network configuration success over monolithic generation [28], and process-observability research shows that execution traces can expose unintended behavior in multi-agent systems [29]. Together, these findings motivate a migration architecture in which source-derived dependency facts precede agent creation and runtime execution.

### D. Testing, Monitoring, and Diagnostic Evidence

Testing of LLM-modified software cannot be reduced to a single generated/not-generated outcome. The migration setting combines at least four validation layers: structural validity of the target SQL or PL/pgSQL artifact, dependency validity against the target schema, execution validity in PostgreSQL, and behavioral validity against source-side expectations. These layers are complementary. A script can be syntactically well formed yet fail because a referenced type, table, function, or package-level construct is unavailable; conversely, a script can execute while still violating transformation intent. This distinction is central to the proposed framework because dependency-aware task construction supplies the context needed before execution, while monitoring and diagnostics classify failures after execution.

Prior peer-reviewed work by one of the authors on evolution-aware specification-driven monitoring [8] treats monitoring specifications as regenerable engineering artifacts that define monitored entities, contexts, functional states, metrics, alerts, strategies, and adaptation rules. Applied to database migration, the same principle suggests that every migration task should carry a compact verification specification: expected object identity, required predecessors, target dialect constraints, executable checks, and recovery policy. This transforms verification from an after-the-fact manual activity into an explicit part of the task contract.

Multi-drift predictive monitoring [9] adds another useful perspective. A migration workflow evolves along several dimensions at once: the generated artifact changes, the target schema grows, available dependencies change between phases, agent prompts or models may vary, and the execution state changes after each successful or failed task. Treating these changes as observable state transitions helps distinguish a local code-generation error from a context or topology error. The distinction matters for diagnosis: a syntax error should normally be repaired locally, whereas an unresolved dependency may require graph reordering, context enrichment, or re-execution of a predecessor rather than another unconstrained generation attempt.

The adaptive monitoring model for LLM-modified information systems [30] further motivates property-oriented validation rather than a single code-coverage indicator. In the present study, the relevant properties include parseability, object preservation, dependency satisfaction, execution success, traceability, token use, retry behavior, and context size. The proposed validation layer therefore does not assume that a high regeneration rate alone means successful migration. Instead, regeneration is treated as one stage of a multi-stage evidence chain whose next decisive gate is executable PostgreSQL validation.

These monitoring-oriented studies are not used as evidence that the current dynamic-agent policy is already optimal. Their role is architectural: they justify explicit state, validation, and feedback signals in the migration pipeline. The empirical evidence for the proposed task decomposition comes from the 116-file decomposition experiments, while the complementary 1,006-file migration study supplies execution-level observations. The controlled comparison of alternative routing policies remains a separate experiment.

TABLE I. POSITIONING OF REPRESENTATIVE APPROACHES

| Approach | Task graph | Dynamic agents | Primary gap for migration |
|---|---|---|---|
| AutoGen [11] | Developer-defined | Configurable | No source-derived DB dependency graph |
| RSTD [14] | Runtime control over fixed subtasks | No | Assumes subtasks already identified |
| TDAG [15] | Dynamic | Yes | Generic tasks, not PL/SQL source analysis |
| VMAO [20] | Planner-generated DAG | Specialized | Query decomposition, not code dependencies |
| AdaptOrch [24] | Input task DAG | Topology-adaptive | Assumes dependency DAG exists |
| Proposed | Derived from Oracle code | Yes | Combines source analysis, DAG, and agent generation |

Overall, the comparison shows that existing orchestration approaches either assume a predefined task graph or operate on generic task representations. The proposed framework differs by deriving the execution graph directly from Oracle source artifacts and using it as the basis for runtime agent generation.

## III. PROBLEM FORMULATION

Let an Oracle migration corpus consist of artifacts $A = \{a_1, ..., a_n\}$. Each artifact is decomposed into migration units. A unit becomes an operational task only when its source fragment, type, dependencies, migration objective, and verification conditions are explicitly represented. We define a task as

$$T_i = < u_i, type_i, dep_i, c_i, r_i, ctx_i, ac_i >, \quad (1)$$

where $u_i$ is the source unit, $c_i$ is complexity, $r_i$ is migration risk, $ctx_i$ is required dependency context, and $ac_i$ is a set of acceptance criteria. This distinction is important because a syntactically valid fragment is not necessarily a migration-ready task.

Dependencies over the task set define a directed graph

$$G = (V, E),$$
$$V = \{T_1, ..., T_m\},$$
$$E \subseteq V \times V, \quad (2)$$

where an edge $(T_i, T_j) \in E$ means that $T_i$ depends on $T_j$ and therefore $T_j$ must be available before $T_i$ is executed or validated. External objects remain explicit graph nodes or external references rather than being silently discarded.

The hybrid extraction rule is deterministic-first:

$$D(u) = \begin{cases} D_{\mathrm{AST}}(u), & \text{if } parse(u) = valid, \\ V(D_{\mathrm{LLM}}(u), u), & \text{otherwise.} \end{cases} \quad (3)$$

where $V$ validates LLM-proposed dependency names against the source fragment and rejects unsupported candidates. This makes the LLM a selective recovery mechanism rather than the primary parser.

Cycles prevent ordinary topological ordering. Let SCC(G) be the strongly connected components identified by Tarjan's algorithm [31]. Condensation maps every SCC to one work item, producing

$$G^* = Condense(G, SCC(G)), \quad (4)$$

which is acyclic. A topological layering then yields phases $P0, ..., Pk$ such that tasks in one phase have no unresolved predecessor relation to one another and can be considered for parallel execution.

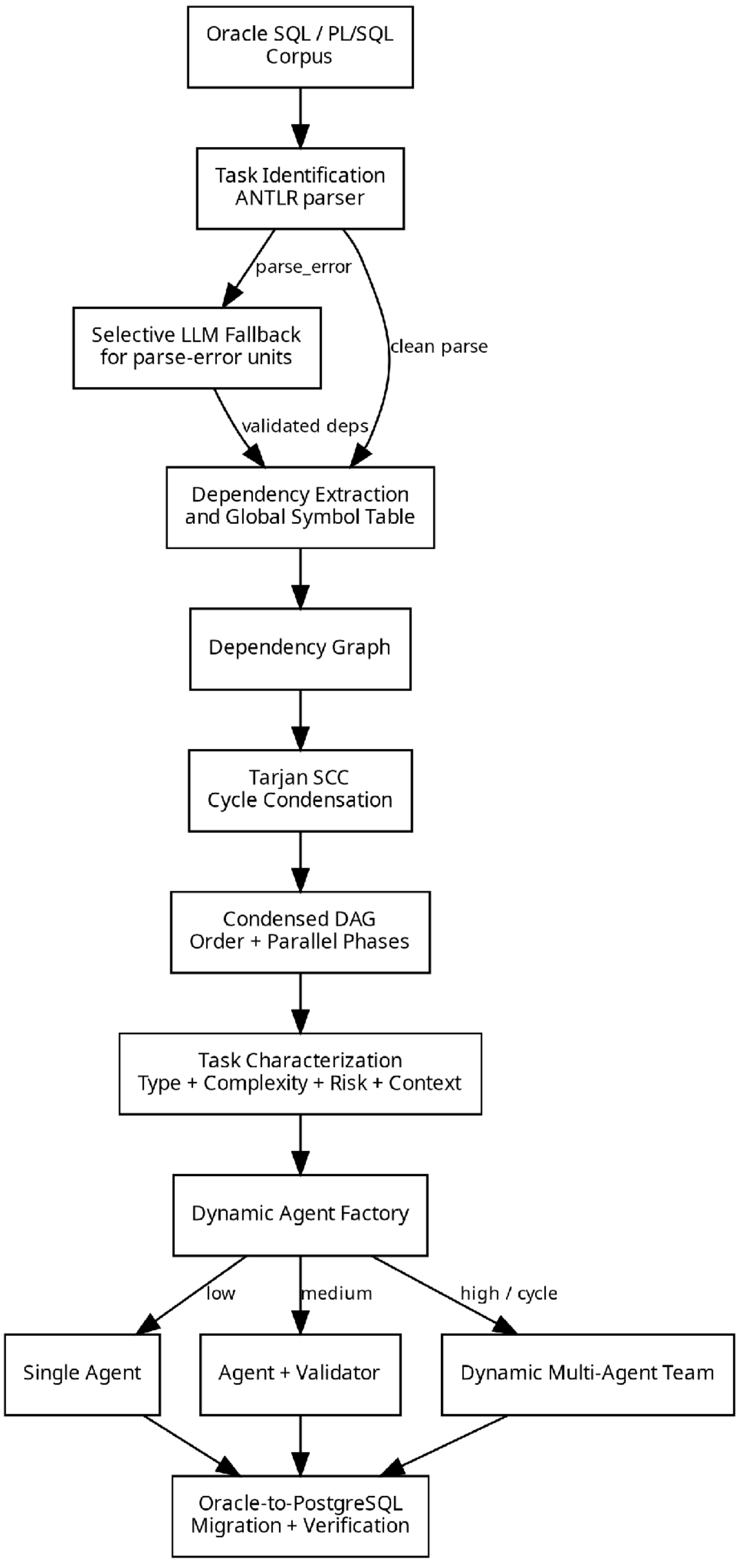

Fig. 1. Proposed hybrid dependency-aware decomposition and dynamic agent generation. PlantUML source is supplied with the manuscript.

Thus, decomposition and agent generation are connected through an explicit dependency-aware task representation rather than through a fixed workflow. This separation allows deterministic source analysis to remain independent from runtime orchestration policy.

## IV. Hybrid Decomposition and Dependency Graph Construction

### A. Deterministic Task Identification

The implementation follows five typed stages: ingest, ANTLR parse, dependency extraction, graph construction, and Tarjan/DAG processing. ANTLR is used because grammar-derived unit boundaries and statement types are more reliable than punctuation-only splitting for mixed Oracle SQL/PL/SQL [32]. The parser identifies procedures, functions, triggers, views, DDL/DML statements, and other top-level units. Coverage checks verify that each source character belongs either to an extracted unit or to an explicitly handled non-unit region.

Earlier splitter experiments clarify why a hybrid design is necessary. A static character scanner is extremely fast but can fail on SQL*Plus/SQLcl scripts that do not consistently use semicolon terminators. On the 116-file corpus, ANTLR provides the high-precision structural path but does not cleanly parse every real-world construct. LLM slicing can remain information-preserving for processed files, but its operational cost and rate limits make it unsuitable as the default path. The architecture therefore assigns each mechanism a narrow role rather than selecting one universal splitter.

TABLE II. Comparison of Oracle Decomposition Engines

| Engine | Corpus behavior | Interpretation |
|---|---|---|
| Static scanner | ~0.7 s; fragile on non-semicolon-terminated scripts | Fast baseline, not a dependable backbone |
| ANTLR | Grammar-derived units; 0 coverage gaps in final pipeline | Primary deterministic structure |
| LLM splitter | 86/86 processed files lossless; 30 rate-limited | Flexible but costly and quota-dependent |
| Hybrid dependency path | ANTLR + selective LLM fallback | Used in the proposed framework |

These results support the deterministic-first design: ANTLR provides the structural backbone, while the LLM is reserved for cases where grammar-based extraction is insufficient. This limits probabilistic processing to the subset of units that actually require semantic recovery.

### B. Dependency Extraction and Global Resolution

A global symbol table maps qualified object names to defining units. References are extracted as typed dependencies, including table/view references, routine calls, sequence access, trigger targets, and DDL targets. Internal references are resolved to task nodes; references not defined in the corpus remain external. Provenance is retained as AST or LLM. Ambiguous resolution is recorded rather than guessed.

The practical implementation improved graph quality by reading statement types directly from the ANTLR parse tree. For example, CREATE TABLE defines an object, while DROP, ALTER, and TRUNCATE reference an existing object and therefore induce ordering dependencies. On the same 116-file corpus, this change increased total dependencies from 1,063 to 1,271 and resolved internal edges from 268 to 446 while preserving zero coverage gaps.

### C. Cycle Handling and Parallel Phases

Tarjan's strongly connected component algorithm runs in linear time O(|V|+|E|) [31]. Every SCC with more than one member, or a self-loop, is represented as a cycle work item before topological sorting. The real 116-file corpus contained no cycles, so cycle behavior was validated separately on a controlled procedure fixture. The original dependency graph in Fig. 2 contains the cycle SVC_A -> SVC_B -> SVC_C -> SVC_A; Fig. 3 shows the condensed DAG where those procedures form a single cycle node.

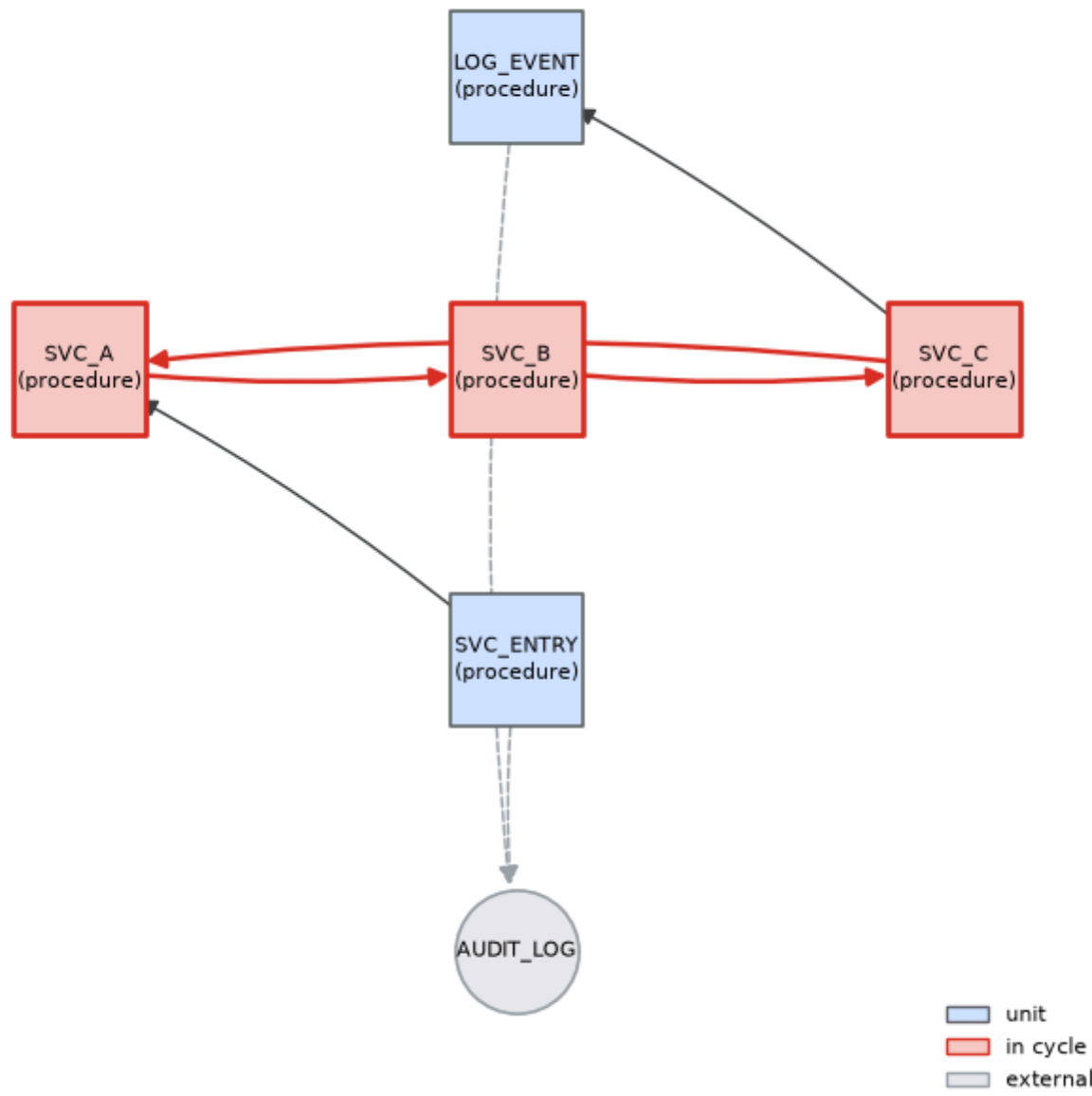

Fig. 2. Controlled cycle fixture before SCC condensation: SVC_A, SVC_B, and SVC_C form a strongly connected component; AUDIT_LOG is external.

The original graph demonstrates why ordinary topological sorting is insufficient when mutually dependent procedures occur. The identified strongly connected component must first be treated as a single dependency unit.

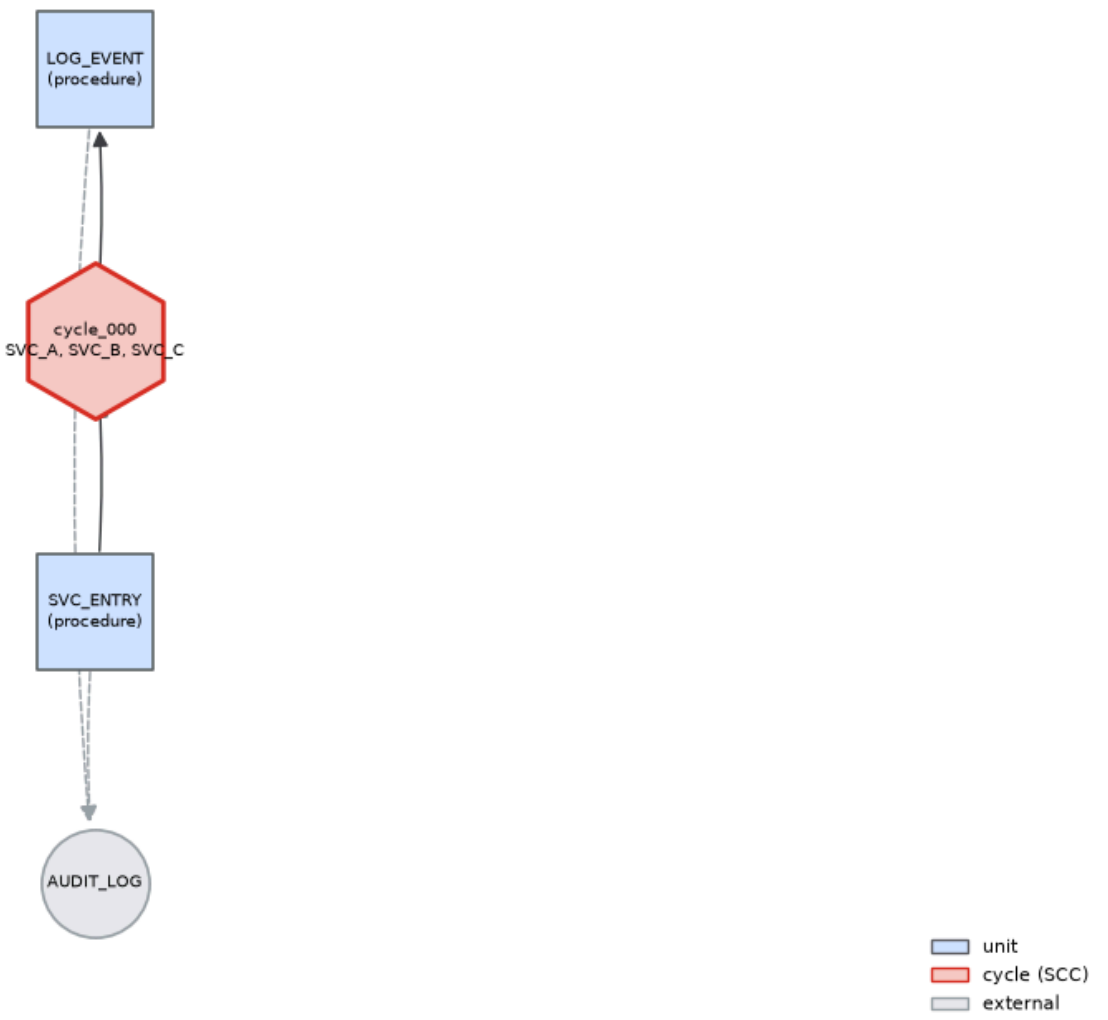

Fig. 3. Condensed DAG after Tarjan SCC processing. The cyclic procedures are represented by one work item while dependency order is preserved.

After SCC condensation, the graph becomes acyclic and can therefore be ordered and partitioned into executable phases. Importantly, the internal cycle is preserved explicitly rather than hidden by an arbitrary ordering of its members.

## V. Dynamic Agent Generation

The dependency DAG is not treated as the final output. Each node or cycle group is converted into a structured migration task and passed to a Dynamic Agent Factory. The factory generates a runtime configuration rather than selecting a permanently hard-coded agent instance. The configuration contains role and instructions, model class and budget, tool permissions, dependency context, output schema, and validation policy.

$$Agent_i = F(type_i, c_i, r_i, ctx_i, ac_i), \quad (5)$$

This formulation makes agent generation conditional on observable task properties and allows the routing policy itself to become an experimental variable. The mapping is intentionally policy-driven and testable. Low-complexity independent units can be executed by a single migration agent. Medium-complexity units use a migration agent followed by a verifier. High-complexity tasks, cycle groups, or tasks with high dependency fan-in can trigger a temporary multi-agent team containing, for example, a migration agent, PostgreSQL specialist, dependency/context agent, and validator. The agents are created for the task and can be discarded after completion, limiting persistent context and communication overhead.

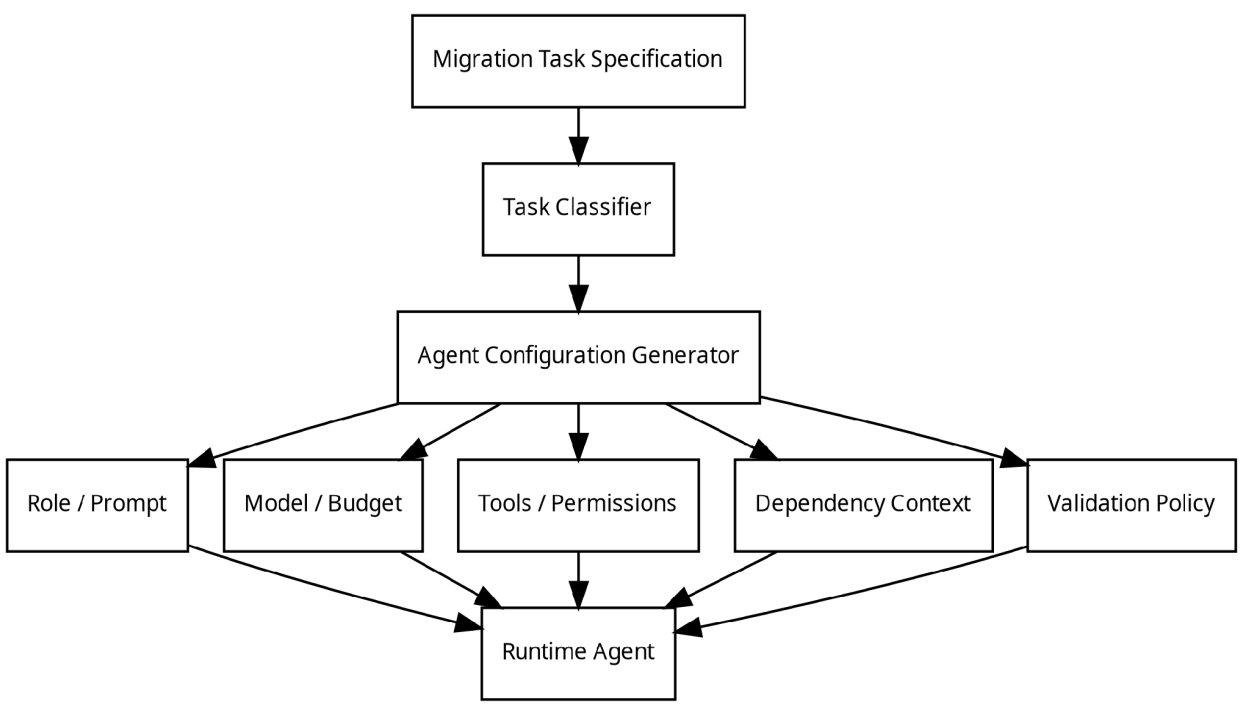

Fig. 4. Dynamic Agent Factory. Task metadata determines the runtime role, model, tools, bounded dependency context, and validation policy. PlantUML source is supplied with the manuscript.

The factory therefore converts structural task metadata into an execution configuration without modifying the source-derived dependency facts. Agent specialization becomes an orchestration decision that can be evaluated independently from decomposition quality.

TABLE III. Proposed Task-To-Agent Routing Policy

| Task class | Runtime topology | Context supplied |
| --- | --- | --- |
| Low complexity | Single migration agent | Task source + direct dependencies |
| Medium complexity | Migration agent + validator | Task + dependency closure + acceptance criteria |
| High complexity | Temporary specialized MAS | Task + selected architecture/schema context |
| Cycle group | Coordinated team / one work item | All SCC members + inbound dependencies |

The policy deliberately increases orchestration complexity only when task characteristics justify it. Low-complexity tasks remain lightweight, whereas dependency-heavy or cyclic tasks receive additional context, validation, or coordination.

## VI. Experimental Design

### A. Experimental Configurations

The completed experiments evaluate decomposition fidelity, dependency recovery, and the execution-level behavior of the specification-first migration baseline. The controlled C0–C3 comparison defined here is designed to isolate migration-level effects that have not yet been measured for dependency-aware dynamic-agent orchestration. Four configurations should be executed on the same Oracle-to-PostgreSQL task set: C0 monolithic migration, C1 functional-unit decomposition with one generic agent, C2 dependency-aware decomposition with one generic agent, and C3 the proposed hybrid decomposition plus dynamic agent generation. This ablation separates gains due to decomposition, dependency ordering, and runtime specialization.

TABLE IV. EXPERIMENTAL CONFIGURATIONS FOR THE C0–C3 ABLATION STUDY

| ID | Decomposition | Dependency order | Agent topology |
|---|---|---|---|
| C0 | None | None | Single monolithic agent |
| C1 | Functional units | No | Static generic agent |
| C2 | Functional units | Yes | Static generic agent |
| C3 | Hybrid selective fallback | Yes + SCC phases | Dynamic by task |

The four configurations provide an incremental ablation path from monolithic processing to the complete proposed framework. Consequently, the effects of decomposition, dependency-aware ordering, and dynamic specialization can be evaluated separately.

The experimental design evaluates this central question at three levels: decomposition fidelity and dependency recovery, construction of a dependency-respecting execution graph, and migration-level effectiveness under alternative orchestration policies. The corresponding evaluation dimensions and hypotheses are specified in Section XV.

### B. Metrics

Source coverage measures whether decomposition preserves the input stream:

*Coverage = preserved source characters /*

*total source characters* (6)

The fallback contribution can be reported as the relative gain in resolved internal dependencies:

$$Gain_{edges} = \frac{E_{hybrid} - E_{AST}}{E_{AST}}. \quad (7)$$

To quantify overhead against a monolithic baseline, we use a decomposition overhead ratio:

$$DOR = \frac{Tokens_{decomp} + sum\ Tokens_{subtasks}}{Tokens_{baseline}}. \quad (8)$$

For end-to-end migration, syntactic validity alone is insufficient. The primary quality vector should include PostgreSQL syntax/compilation validity, dependency satisfaction, executable test pass rate, and semantic/behavioral equivalence where runnable tests exist. A cost-aware effectiveness score can be reported only after those measurements are available:

$$Efficiency = \frac{successful\ verified\ migration\ tasks}{total\ model\ tokens}. \quad (9)$$

The experiment should additionally report wall-clock latency, number of LLM calls, retries, task-level context size, and validation failures. Semantic code-translation evaluation should prioritize executable or behavior-oriented measures over BLEU-like text similarity [33], while CodeBLEU can remain a secondary structural metric [34].

## VII. MEASURED RESULTS OF THE DECOMPOSITION SUBSYSTEM

### A. Corpus-Level Deterministic Baseline

The final deterministic implementation was executed without an LLM on the same 116-file Oracle corpus. All files were processed into 1,037 units with zero coverage gaps. The pipeline extracted 1,271 dependencies, resolved 446 internal edges, and produced 825 external edges. It also generated 44 DDL-target edges after statement roles were derived directly from the grammar. The resulting graph contained four phases. No cycles were observed in this corpus.

The implementation test suite collected 565 tests: 564 passed, while one compatibility assertion failed because of a DDL-versus-OTHER classification mismatch between the grammar-based and legacy regex-based classifiers.

### B. Selective LLM Fallback

The controlled fallback run changed only one variable: parse-error units were routed to the LLM dependency extractor while the AST path remained unchanged. Of the 1,037 units, 165 (15.9%) were parse-error units. The LLM returned candidate dependency names that were validated against the source text with retry support.

The AST contribution remained exactly 1,271 dependencies, while the LLM contributed 496 additional validated dependencies. Consequently, the selective fallback eliminated all 165 dependency-blind parse-error units in the evaluated corpus (165 → 0), and resolved internal edges increased from 446 to 527, corresponding to an 18.2% increase in resolved internal dependencies. Edges not resolved to an internal corpus definition increased from 825 to 1,182; this category includes genuinely external objects as well as references whose internal definitions may remain unresolved.

Sequential fallback processing required approximately 37 minutes, while graph construction and Tarjan processing remained on the order of seconds.

TABLE V. EFFECT OF SELECTIVE LLM FALLBACK ON DEPENDENCY EXTRACTION

| Metric | ANTLR-only | ANTLR + LLM |
|---|---|---|
| AST dependencies | 1,271 | 1,271 |
| LLM dependencies | 0 | 496 |
| Unresolved dependency units | 165 | 0 |
| Resolved internal edges | 446 | 527 |
| External edges | 825 | 1,182 |
| Graph phases | 4 | 4 |
| Coverage gaps | 0 | 0 |

The fallback therefore improves dependency coverage without altering the deterministic AST contribution. The main remaining limitation shifts from dependency extraction to cross-file dependency resolution.

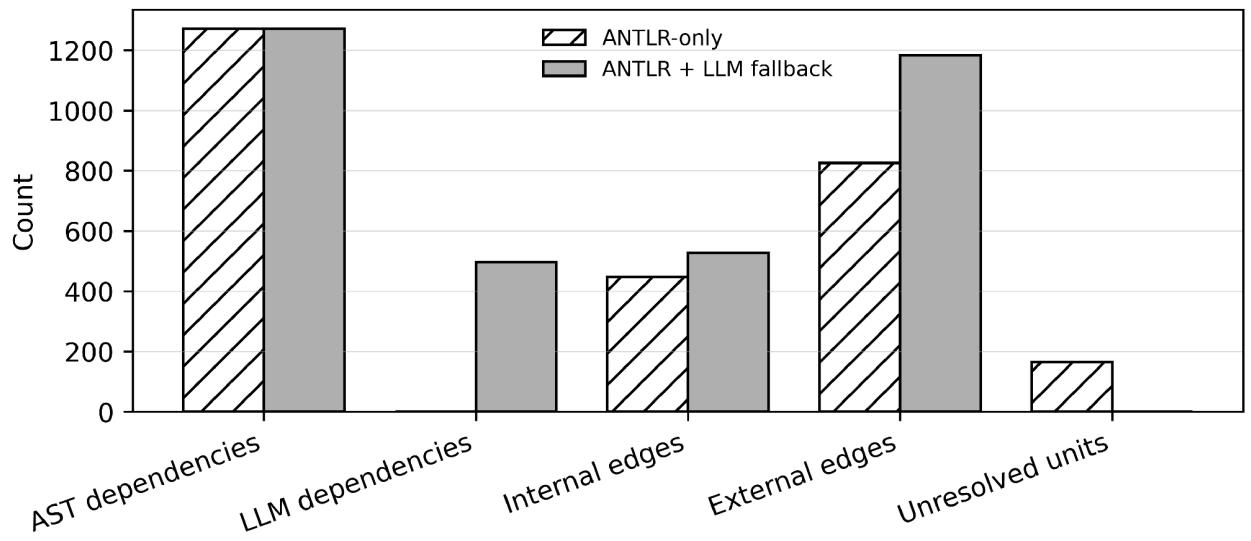

Fig. 5. Effect of selective LLM fallback on dependency extraction and graph completeness. The AST contribution remains unchanged, while the fallback adds validated dependencies and eliminates dependency-blind units.

The comparison highlights that the largest qualitative change is the elimination of dependency-blind units, while resolved internal edges also increase from 446 to 527. At the same time, the growth in unresolved-to-internal references indicates that extracting additional names does not automatically guarantee successful graph resolution.

### C. *Task Complexity Distribution*

The working decomposition pipeline also assigned coarse complexity labels to the 1,037 extracted units: 390 low, 553 medium, and 94 high. These values are not yet evidence that a particular agent topology is optimal; they provide the measurable input distribution for the planned dynamic-agent experiment. The distribution is sufficiently heterogeneous to test whether per-task specialization is more efficient than assigning the same model and tools to every unit.

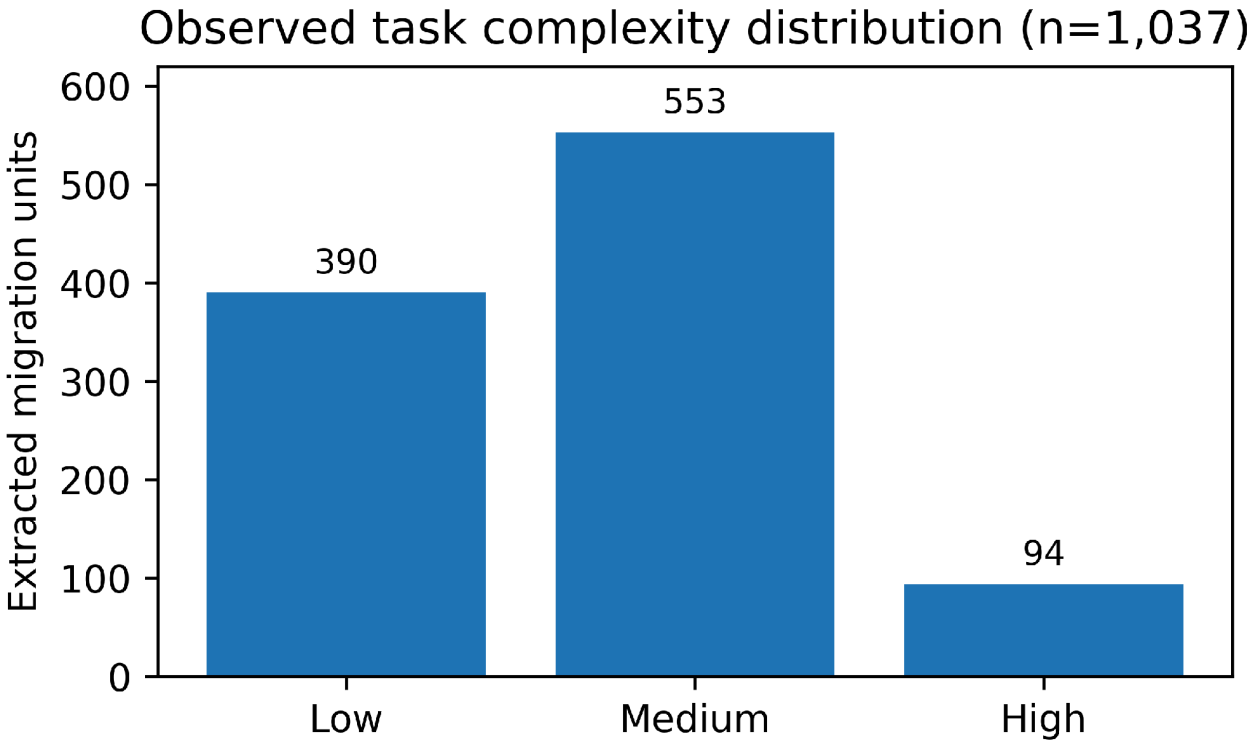

Fig. 6. Observed complexity distribution of the 1,037 extracted Oracle migration units.

The dominance of medium-complexity tasks indicates that the corpus is not concentrated at either extreme of the routing policy. This distribution provides a suitable basis for testing whether differentiated agent configurations outperform a uniform execution strategy.

TABLE VI. COMPLEXITY DISTRIBUTION OF EXTRACTED MIGRATION UNITS

| Complexity | Units | Share |
|---|---|---|
| Low | 390 | 37.6% |
| Medium | 553 | 53.3% |
| High | 94 | 9.1% |
| Total | 1,037 | 100% |

Accordingly, complexity class can be treated as a measurable routing input rather than an assumed property of the migration workload.

## VIII. MIGRATION TASK SPECIFICATION AND PROVENANCE MODEL

### A. *From Parsed Unit to Executable Task Contract*

A source unit becomes useful to an agentic migration system only after it is converted into an explicit task contract. The contract is deliberately richer than a code fragment. It contains the stable unit identifier, source artifact identifier, source object name, object type, original source fragment, source and target dialects, dependencies, dependency provenance, complexity class, migration objective, acceptance criteria, routing hints, and execution state. The purpose of this representation is to separate facts extracted from the source from decisions made by the orchestration layer. A parser can state that a procedure references a table; it should not decide that a particular model must migrate the procedure. Conversely, an agent router can select a model or verifier without redefining which source dependencies exist.

This separation is essential for reproducibility. If an agent output fails, the evaluator must be able to reconstruct the exact task presented to the model. A stable task identifier links the output to the source fragment and to graph state. Dependency provenance records whether a dependency was obtained from the ANTLR parse tree, recovered by the LLM fallback, supplied by a catalog, or corrected by a human. Routing metadata records why a task was classified as low, medium, or high complexity and which contextual artifacts were attached. Together these fields make a migration run inspectable even when the LLM itself is nondeterministic.

The acceptance-criteria field converts the task from an instruction into a testable contract. Criteria should be specific enough to execute automatically whenever possible. For a table, criteria can require successful CREATE TABLE execution, preservation of key columns, compatible types, and expected constraints. For a routine, criteria can require the expected name and signature, successful creation in PostgreSQL, absence of unresolved Oracle-only syntax, availability of required predecessors, and execution of selected tests. Criteria should not contain vague goals such as 'migrate correctly' because such statements cannot drive validation or selective retry.

Routing hints are intentionally advisory rather than authoritative source facts. They can include requires_plpgsql, requires_schema_context, requires_human_review, risk_level, preferred_validator, or cycle_member. The router can recompute or override these hints under a different experiment policy, which is useful when evaluating C2 versus C3. The same task can therefore be executed by a uniform agent in one configuration and by a specialized agent in another without changing the decomposition itself. This design keeps the experimental variable at the orchestration layer rather than confounding it with source parsing.

A cycle group is represented as a composite task rather than as a sequence whose order is invented arbitrarily. Its contract contains the member units and external predecessor set. The internal realization strategy can then be chosen by a cycle-aware policy: joint planning, staged declaration and body creation, temporary stubs, deferred constraints, or explicit human review. The graph layer reports the cycle as a fact; it does not pretend that a topological order exists inside the strongly connected component.

TABLE VII. CORE FIELDS OF AN AGENT-READY MIGRATION TASK

| Field group | Representative fields | Purpose |
|---|---|---|
| Identity | task_id, artifact_id, object_name, unit_type | stable traceability |
| Source | source_fragment, source_dialect, target_dialect | transformation input |
| Graph | dependencies, provenance, phase, cycle_id | ordering and context |
| Assessment | complexity, risk, ambiguity flags | routing policy input |
| Objective | migration_objective, acceptance_criteria | testable task contract |
| Routing | model/tool hints, validator, budget | runtime agent configuration |
| State | status, retries, validation results, artifact hash | observability and recovery |

This task representation separates immutable source-derived evidence from mutable orchestration decisions. The same migration task can therefore be executed under different agent policies without changing its identity, dependencies, or acceptance criteria.

### B. *Dependency Resolution Semantics*

The dependency graph distinguishes extraction from resolution. Extraction asks whether a source unit refers to another object; resolution asks whether that reference can be mapped to a definition inside the migration corpus. These operations have different failure modes. The LLM fallback primarily improves extraction for parse-error units, whereas the large number of external edges after fallback shows that resolution remains the dominant graph-completeness problem. Treating these two stages separately prevents an increase in extracted names from being misreported as an equivalent increase in usable internal ordering information.

Resolution begins with a global symbol table keyed by qualified object name when schema information is known and by a bare name otherwise. Schema qualification follows available source evidence such as DDL prefixes, current-schema statements, or run-level defaults. If one reference maps to one definer, the edge can be treated as resolved. If several definers match, the edge is ambiguous and the alternatives should be retained. If no internal definer is found, the graph records an external node rather than dropping the reference. This conservative representation is important because an 'external' edge can mean a genuinely pre-existing object, a built-in Oracle object, an object outside the provided corpus, or an internal definition missed by the resolver.

Dynamic SQL remains a difficult special case. A reference encoded inside EXECUTE IMMEDIATE may not be visible as a normal grammar-level table or routine reference. Such edges should carry lower confidence or dynamic provenance and should not silently receive the same status as a catalog-confirmed match. The same applies to overloaded routines, package members, synonym resolution, and type dependencies. The graph schema therefore benefits from explicit confidence and ambiguity fields even when the first experiment reports only aggregate internal and external counts.

For migration execution, the direction of an edge is operational: if task Ti depends on Tj, then the validated target artifact for Tj should be available before Ti is executed, unless a cycle-specific strategy overrides the normal rule. This direction permits the DAG to serve simultaneously as an ordering structure and a context-construction structure. The predecessor set determines scheduling; the selected predecessor outputs and contracts determine what context is supplied to the current agent.

### C. *State Model for Agentic Migration*

A task progresses through explicit states rather than disappearing into a single agent call. A minimal state sequence is identified, ready, generated, statically_validated, executable, behaviorally_validated, and accepted, with failure states attached to each transition. The ready state requires that structural decomposition has completed and all mandatory predecessors are either validated or explicitly classified as external contracts. Generated means that a target artifact exists, not that it is correct. Statically_validated means that syntactic and structural checks have passed. Executable means that the target environment accepted the artifact or its test invocation. Accepted is reached only when the task-specific evidence required by its validation profile is satisfied.

This state model helps avoid a common reporting problem in LLM migration experiments: a generated string is sometimes counted as a successful migration even when it was never parsed or executed. The supplied 1,006-file experiment demonstrates why the distinction matters. Regeneration succeeded for 623 files, but only 380 regenerated outputs passed the reported PostgreSQL execution stage. Keeping generation and execution as separate states makes this attrition visible and allows recovery policies to operate on the correct failure class.

State transitions also define what downstream tasks are allowed to consume. A successor should normally receive only predecessor artifacts that reached the validation level required by policy. For example, a generated-but-not-executable function should not be treated as a reliable dependency for a procedure. This validation-gated state transition is analogous to the principle used in runtime-structured agent pipelines: malformed or unvalidated intermediate outputs are prevented from contaminating downstream reasoning.

TABLE VIII. TASK STATES AND EVIDENCE GATES

| State | Required evidence | Downstream use |
|---|---|---|
| Identified | source unit exists | none |
| Ready | dependencies classified / predecessors available | agent execution permitted |
| Generated | target artifact produced | validation only |
| Static-valid | syntax/structure checks pass | optional context by policy |
| Executable | PostgreSQL create/run succeeds | safe predecessor context |
| Behavior-valid | tests/reconciliation pass | strong predecessor evidence |
| Accepted | all task criteria satisfied | final migration artifact |

The state model prevents generated but unvalidated artifacts from being treated as reliable predecessor context. It also provides explicit checkpoints for selective retry, rollback, and downstream dependency activation.

## IX. Implementation, Determinism, and Reproducibility

### A. Typed Five-Stage Pipeline

The implemented decomposition pipeline is deliberately modular: ingest reads input files, parse extracts typed units, extract_deps identifies references, build_graph performs global resolution, and tarjan_dag condenses strongly connected components and derives order and phases. Typed Pydantic models are used between stages. This organization has two advantages for research. First, every stage can be tested independently with fixtures and deterministic expected outputs. Second, one stage can be replaced without redefining the remainder of the experiment. For example, the LLM fallback can be enabled only in extract_deps while leaving parsing, graph construction, and SCC handling unchanged.

The deterministic path is designed to be byte-stable for identical inputs. Stable identifiers, preserved source fragments, explicit dependency records, and deterministic graph algorithms mean that repeated AST-only runs can be compared by diff. This property is useful when evaluating a probabilistic fallback because any change in the deterministic layer would otherwise confound the attribution of recovered dependencies. The controlled fallback experiment benefits directly from this design: the AST dependency count remained fixed at 1,271 while the fallback contributed 496 additional validated dependencies.

Coverage checking is another reproducibility safeguard. After decomposition, every source character should be accounted for by an extracted unit, an intentionally skipped directive/comment region, or another explicitly classified span. A zero-gap result is stronger than simply reporting that a parser returned some units. Earlier static-scanner experiments illustrate the difference: a scanner can emit a plausible list of units while still merging unrelated statements or losing source characters on scripts that use SQL*Plus separators rather than semicolons.

The graph output is serialized separately from the full unit payload. Thin graph nodes contain identity and type, while the complete source fragment remains in the unit store. This avoids duplicating code in graph structures and lets downstream orchestration resolve a node identifier to its full task only when needed. It also supports progressive disclosure: a router can inspect graph metadata before loading potentially large source or predecessor artifacts into the model context.

### B. Regression and Fixture Strategy

The practical implementation is supported by regression fixtures that cover parsing, statement typing, dependency extraction, graph ordering, coverage, visualization, and cycle handling. The supplied implementation report records 565 collected tests, with 564 passing and one compatibility regression associated with a DDL versus OTHER classification mismatch in an older expectation. This result should not be presented as perfect verification; its value is that the remaining incompatibility is explicit and localized rather than hidden by a broad success claim.

A generated dependency fixture containing 8,769 lines is used for broad regression coverage, while small hand-constructed fixtures test specific graph properties. The cycle fixture is especially important because the real 116-file corpus happened to contain no cycles. In that fixture, mutually dependent procedures are correctly collapsed into one SCC work item and the remaining nodes form dependency-respecting phases. This separates algorithmic validation of SCC handling from empirical claims about how frequently cycles occur in the real corpus.

Future reproducibility can be improved further by freezing corpus manifests, parser versions, prompt templates, model identifiers, and validation rules for each reported run. LLM fallback outputs should be cached with provenance so that graph experiments can be rerun without paying the inference cost or introducing new model variability. For externally publishable benchmarks, synthetic or licensed PL/SQL corpora can complement the internal enterprise corpus while preserving the same task and graph schemas.

## X. Operational Scenarios for Dynamic Agent Routing

### A. Low-Complexity Independent Objects

Low-complexity objects with no internal predecessors represent the cheapest execution path. A typical example is a table or sequence whose definition can be transformed without procedural semantics. The router can instantiate one migration agent with a restricted tool set, bounded source context, a target-schema validator, and a small retry budget. If the artifact creates successfully and satisfies structural criteria, no reviewer agent is needed. This scenario tests whether dynamic routing can avoid unnecessary coordination overhead on tasks that the baseline already handles well.

The supplied task distribution contains 390 low-complexity units, making this class large enough to evaluate cost discipline. A successful dynamic policy should not merely improve difficult tasks; it should also preserve or reduce the cost of easy tasks by resisting the temptation to assign every task a multi-agent team.

### B. Dependency-Heavy Procedural Objects

Procedures and functions are the primary scenario for dependency-scoped context. Before generation, the router gathers validated predecessor interfaces and target artifacts, along with the source specification for the current routine. A PL/pgSQL-specialized agent generates the candidate, and an execution validator attempts creation and selected invocations. Missing-object failures are diagnosed against the graph: if the referenced object is an unresolved internal candidate, the resolver is revisited; if it is a later task, scheduling is corrected; if it is truly external, its contract is added to context or the task is blocked for operator input.

This scenario directly addresses the failure pattern observed in the 1,006-file baseline, where functions and procedures frequently failed when the underlying schema did not match. The proposed system does not assume that dependency context will solve every procedural transformation problem. Oracle exception semantics, packages, autonomous transactions, cursors, and dialect-specific built-ins still require transformation logic. The experiment asks a narrower question: how much of the observed failure surface is attributable to missing or incorrectly ordered context, and how much remains after that source of error is controlled?

### C. Cycles, Ambiguity, and Human Escalation

Cycle groups require a different execution policy because no member can simply wait for every internal predecessor to be fully accepted. The orchestrator can create a temporary coordinated agent task that sees all members of the SCC and chooses an explicit cycle-breaking strategy appropriate to the object types. For mutually recursive routines this may involve compatible declarations or staged creation; for circular foreign keys it can involve creating tables before adding constraints. The important point is that the exception is visible and localized to the SCC rather than silently violating global dependency order.

Ambiguous symbol resolution is another escalation trigger. When a bare reference matches several possible definers, an LLM should not be allowed to guess silently because the resulting edge changes both ordering and context. The preferred policy is to preserve all candidates, attempt schema-based disambiguation, consult catalog metadata if available, and request human confirmation when the ambiguity remains material. Such cases are useful for measuring the practical boundary between automated graph construction and expert intervention.

Human-in-the-loop review should therefore be risk-based rather than universal. High-complexity tasks, unresolved ambiguities, repeated execution failures, security-sensitive transformations, and cycle strategies can trigger review. Routine low-risk objects that satisfy all automated gates should proceed without manual approval. This keeps human effort focused on cases where the available evidence does not justify autonomous continuation.

### D. Alternative Route for Query Tasks

The zero-success query result in the supplied specification-first baseline is an important negative finding rather than a reason to discard query migration. It suggests that the chosen spec-write-code loop was poorly matched to that class. Queries often have no persistent target object and can be evaluated more directly through parseability, result equivalence, and performance-oriented checks. A dynamic router should therefore be able to choose a direct dialect-translation strategy for eligible query tasks while retaining the specification-driven path for objects where traceability and structural generation add value.

This alternative route is scientifically useful because it prevents the proposed architecture from being defined as 'dynamic' only in name. If every task ultimately receives the same generation pattern, then complexity-aware routing has little operational meaning. Allowing object classes to select different migration strategies creates a stronger experiment: C3 can be evaluated not only on model specialization but on whether the router chooses an appropriate execution pattern under a common verification framework.

## XI. Failure Containment, Security, and Deployment Considerations

### A. Failure Containment Across the Dependency Graph

Dependency-aware orchestration changes the blast radius of a failure. In a monolithic migration attempt, one malformed output can invalidate the whole script and make it difficult to identify which source object or transformation decision caused the problem. In the proposed graph, every node has a bounded source fragment, explicit predecessors, and a validation state. A failed node can therefore be quarantined while independent nodes in the same or later-unblocked phases continue. The orchestration layer should propagate failure only along true dependency edges, not across unrelated tasks. This is particularly important for large schemas where a small number of difficult packages or routines should not prevent independent tables, views, or sequences from being migrated and validated.

Failure containment also requires immutable checkpoints. Once a predecessor artifact passes its required validation gate, later retries should reuse that artifact rather than regenerate it unless evidence specifically invalidates the predecessor. Reusing validated state reduces token cost and prevents an unrelated retry from introducing new drift into already accepted work. If a predecessor is later superseded, the graph can identify exactly which descendants depend on it and therefore require revalidation. This provides a deterministic alternative to broad full-pipeline reruns.

The same mechanism supports rollback. Because every accepted artifact is linked to a task identifier, specification, dependency set, model configuration, and validation record, the system can restore the last accepted version of one task without discarding unrelated outputs. For experimental evaluation, rollback events should be logged separately from model retries because they reflect a change in validated state rather than a simple inference failure.

### B. Security and Tool Boundaries

Dynamic agent generation must not imply dynamic privilege expansion. Tool access should be derived from a fixed registry and constrained by task role. A migration agent that writes candidate SQL does not need arbitrary filesystem or production-database privileges; an execution validator can be restricted to an isolated PostgreSQL container; a schema-inspection tool can be read-only. High-risk actions such as modifying authentication objects, external links, or privileged routines should require an explicit policy gate or human approval. These boundaries are consistent with earlier work on risks in LLM-based software development [10], where tool misuse, hallucinated actions, and weak traceability are treated as distinct operational risks.

The task specification should therefore include both functional acceptance criteria and execution constraints. A generated artifact can be correct with respect to SQL semantics and still violate migration policy by creating an object in the wrong schema, using disallowed extensions, or embedding credentials. Validation should check such constraints before execution whenever possible. The agent should receive only the tools and credentials needed for the current task, and secrets should remain outside the prompt context. This design makes agent specialization compatible with least-privilege engineering rather than turning specialization into unrestricted autonomy.

Security monitoring can be integrated into the same event stream used for migration diagnostics. Unexpected tool calls, attempts to access unregistered resources, repeated policy-gate failures, and generated statements outside the task's object scope can be recorded as policy deviations. These events are different from translation errors and should trigger different recovery actions, typically blocking execution and escalating rather than asking the same agent to retry freely.

### C. Deployment-Oriented Interpretation of Efficiency

Efficiency in a production migration should be interpreted as verified progress per unit of cost, not as raw model speed. A configuration that produces many scripts quickly but leaves a large execution-repair burden can be less efficient than a slower configuration that yields more executable artifacts on the first pass. The supplied baseline illustrates this distinction: regeneration reached approximately 62% of source files, yet only about 61% of regenerated scripts passed the PostgreSQL execution stage. The gap represents downstream work that a generation-only metric would miss. For this reason, the planned C0-C3 comparison should report at least three denominators: cost per generated artifact, cost per executable artifact, and cost per accepted artifact after all required validation. Retry cost should be attributed to the task that caused it, while shared graph-construction cost can be amortized across the run. This accounting makes it possible to determine whether a more sophisticated dependency-aware or multi-agent policy earns back its overhead by reducing failed execution and repeated context ingestion.

A final deployment consideration is incremental migration. Enterprise databases are rarely transformed in a single immutable batch. New files, changed routines, or corrected schema metadata can arrive after an initial run. Stable task identifiers and dependency-aware invalidation make incremental recomputation possible: only changed nodes and affected descendants need to be reprocessed. Although incremental-update performance is outside the current experiment, the architecture deliberately preserves the information required for that extension.

## XII. End-to-End Migration Generation and PostgreSQL Validation

### A. Specification-First Baseline Pipeline

The decomposition study establishes that a migration corpus can be transformed into a lossless dependency-aware task graph, but graph quality alone does not demonstrate that generated PostgreSQL artifacts are executable. To add an execution-level baseline, we incorporate the supplied specification-first migration experiment. Its purpose differs from the 116-file graph experiment: the graph experiment measures structural decomposition and dependency recovery, whereas the migration experiment measures whether a complete generation pipeline can regenerate PL/SQL artifacts and execute the resulting scripts in PostgreSQL 16.

The baseline pipeline begins with source SQL, removes non-essential SQL*Plus directives and other lines that need not be sent to the model, extracts source objects, and then invokes an agent loop for each object. The loop first writes a specification and then generates target code from that specification. This separation is important for traceability: the specification records how the source object was interpreted and provides an intermediate artifact that can be inspected or regenerated independently of the final SQL/PL/pgSQL output. The approach also localizes rollback because one problematic object can be rerun without repeating every object in the source file.

This baseline already contains two ideas that are retained in the proposed framework: object-level isolation and explicit intermediate specification. The proposed method extends them with dependency-aware ordering, cross-file context, complexity-aware agent creation, and validation-driven recovery. In other words, the baseline demonstrates the feasibility of a specification-first generation loop, while the new framework addresses the main limitation exposed by its failures: the agent does not always know which previously migrated objects, schema facts, or predecessor outputs must be visible when the current object is generated and executed.

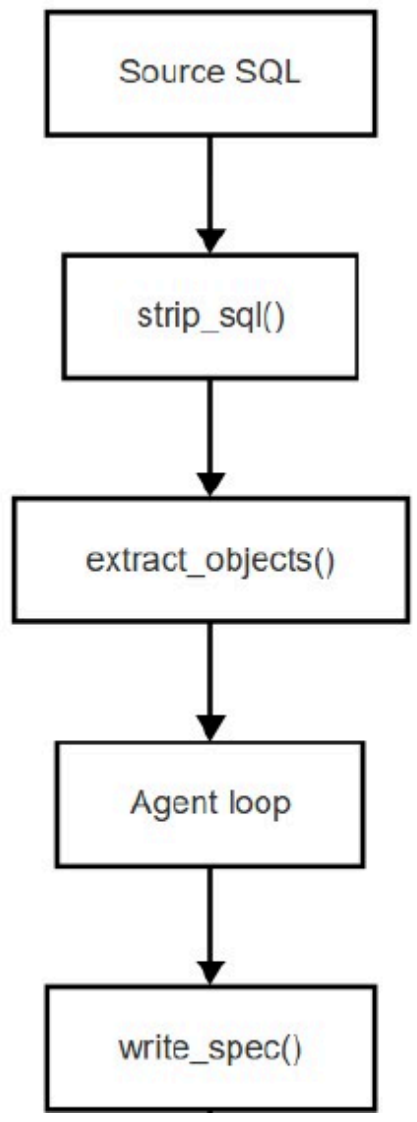

Fig. 7. Specification-first migration baseline used in the supplied end-to-end experiment: source SQL is stripped and decomposed before per-object specification and code generation.

This baseline isolates migration objects before generation but does not yet use cross-file dependency ordering or dependency-scoped context. It therefore provides a useful execution-level reference point for the proposed dependency-aware extension.

### B. Agent Loop and Termination Behavior

The supplied implementation executes each extracted object through an iterative agent loop. The model receives messages together with tool schemas and can call write_spec, write_code, or finish through an MCP server. The loop is capped at ten iterations, although the reported runs averaged approximately three iterations. A successful explicit finish tool call returns a finished outcome. If the model stops without a tool call, the implementation records a no_tools outcome; if the loop exceeds its bound without a terminal outcome, it is marked exceeded. This control structure is operationally relevant because a migration result can fail before SQL validation even begins: an agent can terminate incorrectly, fail to invoke the expected tool, or spend several iterations without producing a usable artifact.

Two models were used in the baseline generation experiment: gpt-oss-120b and gpt-o4-mini. The supplied results report that gpt-oss-120b was faster on average, while token consumption was broadly similar across the two models. A concrete orchestration issue was observed with gpt-oss-120b: it did not reliably call the finish tool and therefore some runs ended with the no_tools flag. This behavior did not change the reported migration-performance counts, but it is important for production orchestration because a control-plane failure should be distinguished from an SQL-generation failure. The dynamic-agent design therefore treats termination policy and tool-call compliance as monitorable properties of an agent configuration, not as incidental implementation details.

Average token consumption was reported at approximately 2,000 tokens per iteration, with about three iterations per object on average. The observed range depended strongly on object size: some small tables used roughly 900 tokens, whereas larger objects reached approximately 3,000 tokens. These measurements support the context-bounded design of the proposed method. If a task receives the whole corpus rather than a dependency closure, token cost scales with irrelevant context as well as required context. Dependency-aware progressive disclosure gives the orchestration layer a deterministic basis for limiting context while retaining the source objects that the task actually depends on.

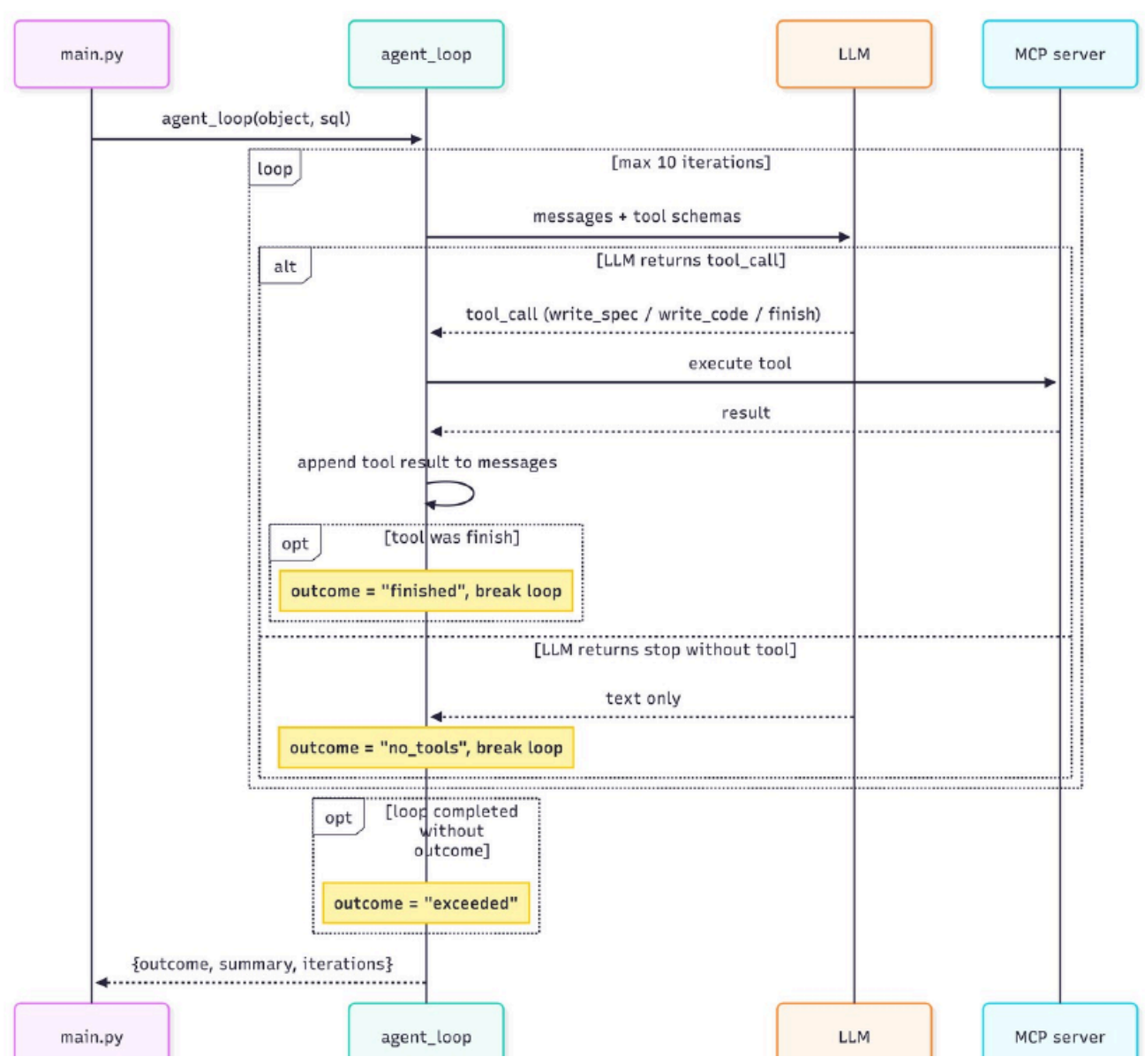

Fig. 8. Agent-loop sequence used by the baseline migration implementation, including explicit finish, no_tools, and iteration-limit outcomes.

The sequence also shows that orchestration failure and migration-code failure are distinct events. Explicit termination states make it possible to diagnose agent-control problems separately from SQL translation errors.

## C. Measured End-to-End Results

The migration baseline was executed on 1,006 PL/SQL files. Of these, 623 files, approximately 62%, underwent successful regeneration. This is a generation-stage result and should not be confused with executable migration success. The generated scripts were subsequently executed in a PostgreSQL 16 Docker environment, where 380 of the 623 regenerated scripts, approximately 61% of regenerated outputs, executed successfully. Relative to all 1,006 source files, this yields a derived end-to-end source-to-execution rate of 37.8% (380/1,006), calculated from the two reported stages as a funnel indicator rather than as an independently measured migration-quality metric.

Performance differed substantially by object class. Tables were the most successful objects, at approximately 85% regeneration success, with views and triggers reported as the next strongest categories. Queries were the weakest category: no query regeneration succeeded in that experiment, leading the original study to suggest that direct translation could be more appropriate for queries than specification-mediated regeneration. User-defined functions and procedures also required special developer attention. During PostgreSQL execution, functions, procedures, and indexes failed most often because the required underlying schema did not match the assumptions of the generated artifact.

This object-level pattern is particularly relevant to the current dependency-aware framework. Tables are often relatively self-contained definitional objects, whereas functions, procedures, indexes, triggers, and views can depend on a target schema that is only partially materialized at generation time. A flat per-object pipeline therefore risks treating a context failure as a generation failure. The dependency graph provides a mechanism to test this hypothesis directly: if a procedural task receives a verified dependency closure and is scheduled after the required predecessors, the change in execution success can be measured without altering the source object itself.

The results also motivate a differentiated task policy. A class that already migrates reliably may not justify the cost of a complex multi-agent team, while dependency-heavy procedural objects may benefit from stronger context, a specialized PL/pgSQL role, an execution validator, and a retry policy that preserves successful predecessors. This is the central reason dynamic agent generation is framed as task-conditioned orchestration rather than as multi-agent execution by default.

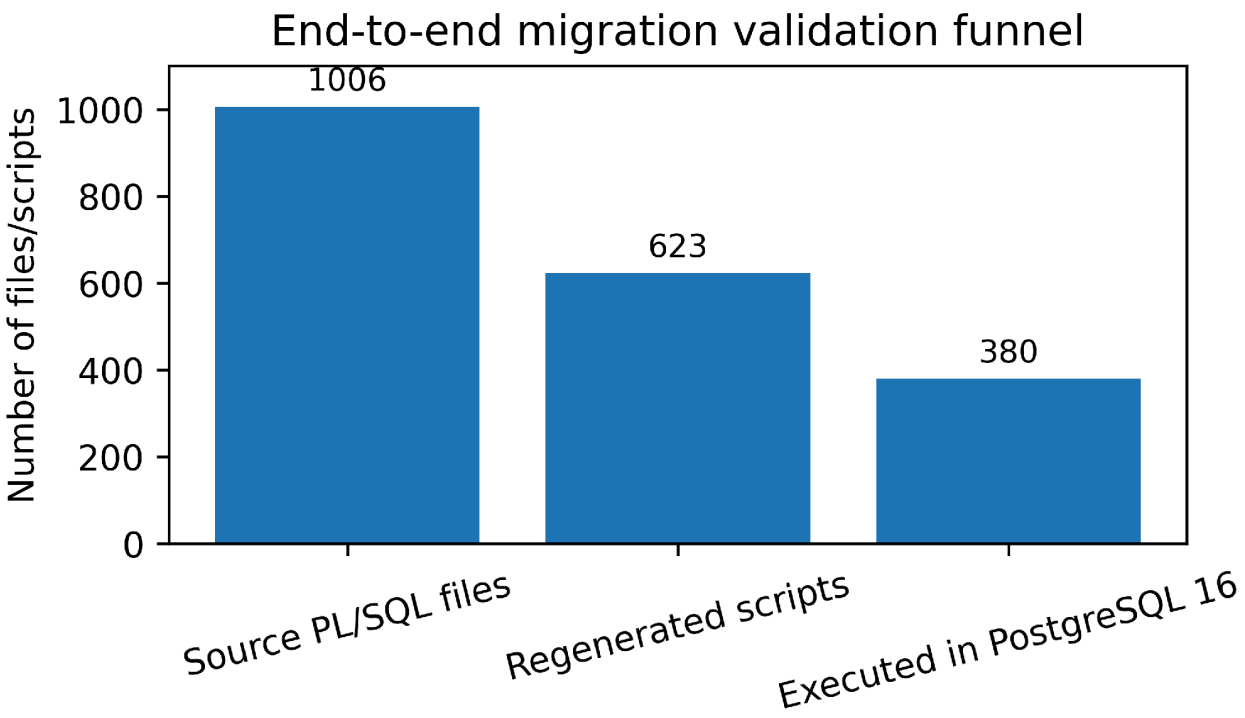

Fig. 9. End-to-end validation funnel from the supplied migration experiment: 1,006 source PL/SQL files, 623 regenerated scripts, and 380 scripts executing successfully in PostgreSQL 16.

The funnel exposes substantial attrition between source ingestion, regeneration, and executable target artifacts. In particular, successful code generation should not be interpreted as successful migration without target-side execution evidence.

TABLE IX. End-To-End Migration Baseline Results

| Measure | Observed result | Interpretation |
|---|---|---|
| Source files | 1,006 | PL/SQL inputs tested |
| Successfully regenerated | 623 (~62%) | Generation-stage success |
| Successfully executed in PostgreSQL 16 | 380 / 623 (~61%) | Execution success among regenerated outputs |
| Overall source-to-execution | 380 / 1,006 (37.8%) | Derived funnel rate |
| Tables | ~85% regeneration | Strongest reported object class |
| Queries | 0% regeneration in the evaluated specification-first baseline | Weakest reported class |
| Avg. iterations | ~3 | Agent-loop average |
| Avg. tokens / iteration | ~2,000 | Input/output-size dependent |
| Observed token range | ~900-3,000 | Reported for small to larger objects |

These results motivate the multi-level validation architecture introduced next: the baseline demonstrates both that corpus-scale generation is feasible and that generation success alone leaves a substantial unresolved execution gap.

## XIII. Testing, Monitoring, and Diagnostic Layer

### A. Why Execution Testing Must Be Multi-Level

A migration pipeline has several distinct failure modes that become conflated if validation is represented by one binary flag. The first level is decomposition validity: every relevant source character and source object must be represented by a migration unit. The second level is generation validity: the agent must produce the required target artifact and satisfy the task output schema. The third level is static target validity: PostgreSQL syntax, object names, signatures, type mappings, and forbidden Oracle constructs must be checked. The fourth level is dependency validity: referenced objects must exist, be ordered correctly, and expose compatible interfaces. The fifth level is execution validity in a real PostgreSQL environment. The sixth level, when test fixtures or source behavior are available, is behavioral equivalence. A useful evaluation must state which of these levels it measures instead of labeling all of them simply as migration success. Classical software-testing research shows that coverage or successful execution alone does not establish test adequacy or fault-detection effectiveness [35], [36].

The proposed framework therefore associates every task with acceptance criteria and a validation profile. For a table, the profile can include parseability, object creation, column/type checks, key and constraint preservation, and optional row-level reconciliation. For a view, it additionally includes dependency availability and query execution. For a procedure or function, the profile extends to callable signature checks, successful PL/pgSQL creation, dependent-object resolution, and behavior on selected test inputs. For an index, successful creation must be interpreted in the context of the table and column definitions it depends on. This task-specific validation is more informative than a global script pass/fail outcome.

The validation layer is also the main feedback channel for dynamic orchestration. A validation failure is classified before the system decides whether to retry. Syntax and local translation errors normally trigger a targeted regeneration of the current task. Missing predecessor objects or incompatible schema state trigger context enrichment or a graph-level scheduling correction. Tool-control failures such as no_tools trigger an agent-policy repair rather than SQL rewriting. Repeated failures on a high-complexity task can escalate from a single agent to an agent-validator pair or a temporary multi-agent team. Human review remains an explicit option for unresolved high-risk cases.

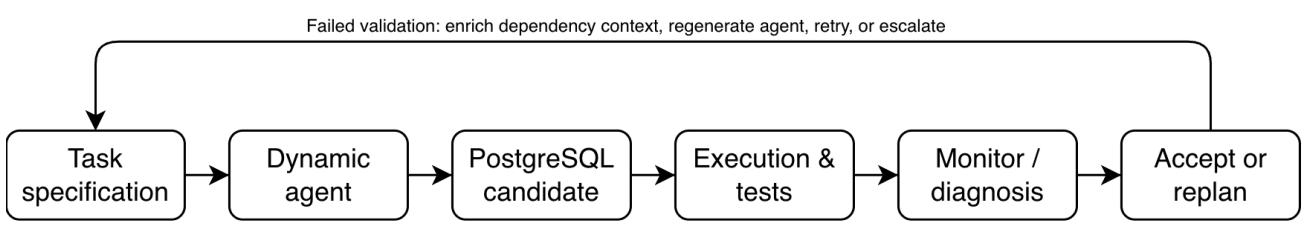

Fig. 10. Validation-driven execution loop. Failed checks are diagnosed before retry so that syntax repair, dependency-context enrichment, agent regeneration, and escalation remain distinguishable recovery actions.

The recovery path is therefore selected from the diagnosed failure class rather than from a generic retry rule. This makes retries observable and limits unnecessary regeneration of already validated dependencies.

### B. Task-Level Observability

Task decomposition creates natural observability boundaries. Each task can emit an execution record containing its task identifier, source object, phase, predecessor set, agent configuration, model, tool calls, prompt/context size, generated artifact hash, validation outcomes, retry count, latency, and token usage. The record should also preserve dependency provenance: whether an edge came from the AST, the LLM fallback, an authoritative catalog, or an operator correction. This allows a failed migration to be traced back not only to an agent output but also to the graph facts that determined what context the agent received.

The monitoring role is not merely operational telemetry. It supports scientific evaluation because the same event structure can be used to compare configurations C0-C3 at task granularity. For example, token efficiency can be conditioned on object type, complexity, dependency degree, or validation outcome rather than averaged across the entire corpus. Retry behavior can be separated into local generation repair, dependency repair, and orchestration repair. Such stratification is necessary because a lower global token count can hide a large cost concentration in a small class of complex procedural tasks.

The monitoring perspective is also consistent with process-observability work for agentic systems [29], which treats event traces and control-flow variation as first-class evidence. In the migration setting, expected variability should come from task type, graph phase, complexity, and validation feedback. Unintended variability includes role violations, unsupported tool calls, repeated generation without state progress, and diverging routes for equivalent tasks. Recording these differences gives the evaluator a way to distinguish deliberate dynamic routing from unstable agent behavior.

### C. Diagnostic Categories and Recovery Actions

For the experiments, failures should be classified into categories that map to specific corrective actions. A parsing/decomposition failure indicates that task identification is incomplete and should activate the LLM fallback or manual parser extension. A dependency-extraction failure means the source unit is known but its references are not; this is the blind spot already measured by units_deps_failed. A dependency-resolution failure means a reference was extracted but could not be linked to a defining unit. A generation failure means the agent did not produce the required artifact. A syntactic validation failure means target code cannot be parsed or created. An execution failure means the artifact is syntactically acceptable but fails against the actual PostgreSQL schema or runtime. A semantic or behavioral failure means execution succeeds but the result differs from expected behavior.

These categories make selective recovery testable. The recovery action should be the narrowest action capable of addressing the diagnosed problem. Re-running the entire pipeline after a missing type or table is both expensive and scientifically uninformative. Instead, the orchestrator should identify whether the missing dependency is external, unresolved, or simply scheduled later; then it can enrich context, postpone the task, or request human resolution. This design follows the same operational logic that makes runtime-structured decomposition effective in other software-engineering workloads [14], while adapting it to migration-specific dependency evidence.

TABLE X. DIAGNOSTIC CLASSES AND SELECTIVE RECOVERY

| Failure class | Primary signal | Preferred recovery |
|---|---|---|
| Decomposition | coverage gap / parse error | parser fallback or split repair |
| Dependency extraction | unit has no validated refs | LLM fallback / rule extension |
| Dependency resolution | reference has no internal target | schema qualification / catalog / context repair |
| Agent control | no_tools / exceeded loop | regenerate agent policy or tool contract |
| Target syntax | PostgreSQL parse/create failure | local code regeneration |
| Target execution | runtime/schema error | dependency context + schedule correction |
| Behavioral | tests/reconciliation differ | semantic repair or human review |

Mapping each diagnostic class to a specific recovery action converts monitoring signals into explicit orchestration decisions. This relationship is required for the dynamic-agent policy to adapt execution without indiscriminate full-pipeline retries.

# XIV. DYNAMIC AGENT GENERATION AS A TASK-CONDITIONED POLICY

## A. Agent Specification

Dynamic generation in this study does not mean synthesizing an unconstrained autonomous persona. It means compiling a task specification into an executable agent configuration. The configuration contains a role, model class, system instruction, permitted tools, dependency context, maximum iteration count, token budget, output schema, validator set, and escalation policy. All fields are derived from task facts or deployment policy. This keeps the dynamic layer auditable: two tasks with the same policy-relevant features should receive equivalent configurations even if their source code differs.

The task-to-agent mapping uses information already available from decomposition. Unit type determines the core migration skill; complexity controls how much reasoning or review is justified; dependency closure determines which predecessor artifacts are placed into context; risk controls the strength of validation; and acceptance criteria define the stopping condition. Cycle membership is also relevant because an SCC has no valid internal topological order. A cycle group can therefore be treated as one coordinated work item, potentially using a joint plan or staged stub-and-fill strategy chosen by a cycle-aware agent policy rather than by independent agents that each assume their dependencies already exist.

The dynamic factory should remain conservative about model choice. A low-complexity table with no unresolved dependencies may be routed to a cheaper model with a syntax validator. A medium-complexity view or trigger can use a migration agent plus a validator. A high-complexity package-derived procedure or cycle group may justify a stronger reasoning model and a specialized reviewer. This policy is a hypothesis to test, not a claim that larger teams are always better. Coordination has cost, so the experiment must measure whether specialization gains exceed additional token and latency overhead.

## B. Dependency-Scoped Context Construction

A central advantage of the DAG is that it provides a principled context-selection mechanism. Let $Pred(T_i) = \{T_j | (T_i, T_j) \in E\}$ denote the direct predecessors of task $T_i$ and $Anc(T_i)$ its transitive dependency closure. The simplest policy supplies the source specification for Ti plus validated outputs for $Pred(T_i)$. A broader policy supplies the minimal subset of $Anc(T_i)$ required to resolve referenced symbols. External objects are represented as contracts or schema metadata rather than as generated predecessor outputs. This avoids the two extremes of context starvation and whole-corpus prompting.

Context construction should preserve provenance. A generated PostgreSQL predecessor is not equivalent to its original Oracle definition, and a schema-catalog fact is not equivalent to an LLM-inferred edge. The context bundle should therefore identify each item as source code, generated target code, schema metadata, specification, validation result, or external contract. Such typing reduces accidental use of stale or inappropriate information and makes the migration trace auditable after completion.

The 1006-file baseline gives a concrete reason to evaluate this mechanism. Functions, procedures, and indexes were among the object types that failed most often during PostgreSQL execution because the underlying schema needed to match exactly. A dependency-scoped context does not guarantee correctness, but it directly targets this failure mode by supplying the agent with verified predecessor artifacts and by deferring execution until required predecessors are present.

## C. Runtime Adaptation and Selective Retry

Runtime adaptation begins after a validation result, not before it. If a task passes all required gates, its target artifact becomes validated state and is available to successor tasks. If it fails, the orchestrator records the failure class and chooses a recovery action. A local syntax error can be retried with the same dependency context but a repair prompt. A missing dependency can trigger a graph lookup and context update. An ambiguous dependency can block the task until a disambiguation rule or human decision is available. Repeated failures can trigger model escalation or a verifier. This sequence keeps retry semantics tied to evidence rather than to a generic self-correction loop.

This design also limits token bleed. RSTD results in a different software-engineering domain show that static decomposition can increase retry cost when downstream subtasks must be re-executed, whereas subtask-level recovery reduces reprocessing [14]. The migration graph provides the state needed to apply a similar principle: once a predecessor artifact is validated, an unrelated local failure should not cause it to be regenerated. The proposed experiments therefore record retry tokens separately from baseline execution tokens and distinguish retries that reuse validated predecessors from full-pipeline restarts.

## XV. Extended Evaluation Protocol

### A. Two Experimental Corpora and Their Roles

The study uses two complementary experimental datasets, and their roles must remain separate. The first is the 116-file corpus used for decomposition, dependency extraction, graph construction, and LLM fallback evaluation. It supplies the strongest controlled evidence because the AST layer was held constant when the fallback was enabled. The second is the 1,006-file PL/SQL migration dataset used for specification-first regeneration and PostgreSQL execution. It supplies a larger execution-level baseline but does not isolate the effect of dependency-aware dynamic agents. Treating the two datasets as one would overstate what has been measured; keeping them separate allows each to support the claim it actually tests.

The 116-file corpus provides evidence for decomposition fidelity, dependency recovery, and graph construction, whereas the controlled C0–C3 migration study is required to evaluate migration-level effects of dependency-aware context and task-conditioned agent generation. The definitive controlled C0–C3 experiment should re-run a matched subset of migration tasks using the same source objects, target environment, acceptance criteria, and repeated runs where stochasticity can affect the outcome.

TABLE XI. Experimental Evidence Layers

| Evidence layer | Dataset | Evaluation role | Measured outputs |
|---|---|---|---|
| Decompositio n / graph | 116 Oracle files | Decomposition and graph validation | units, coverage, deps, edges, phases, fallback |
| Migration baseline | 1,006 PL/SQL files | Execution-level baseline | regeneration, PostgreSQL execution, tokens, iterations |
| Controlled C0-C3 study | matched subset (planned) | Dependency-context and routing ablation | quality, retries, cost, latency, context, execution |

The separation of evidence layers prevents results from the two existing corpora from being conflated with the planned C0–C3 comparison and ensures that each dataset supports only the evaluation role permitted by its experimental design.

### B. Evaluation Dimensions and Hypotheses

The central research question is operationalized through five evaluation dimensions. First, decomposition fidelity measures whether source content is preserved when migration units are identified. Second, fallback recovery measures the additional dependency evidence obtained from parse-error units without changing the deterministic AST path. Third, graph validity evaluates whether the extracted dependencies can be transformed into a cycle-aware, dependency-respecting execution structure. Fourth, migration effectiveness measures the impact of dependency-aware task construction on executable PostgreSQL outcomes. Fifth, orchestration efficiency evaluates whether task-conditioned dynamic agent generation improves verified migration quality relative to token, latency, and retry cost.

The corresponding hypotheses are defined as follows.

H1: Hybrid decomposition preserves source coverage while producing explicit migration units.

H2: Selective LLM fallback reduces dependency-blind units without changing the deterministic AST contribution.

H3: SCC condensation produces a valid dependency-respecting execution DAG for cyclic task structures.

H4: Dependency-scoped task context increases executable migration success for objects with cross-object dependencies.

H5: Task-conditioned agent routing improves verified migration quality per token/cost relative to a uniform agent topology.

### C. Evaluation Metrics and Statistical Reporting

The primary migration outcome should be executable task success: a generated artifact passes the task-specific PostgreSQL creation or execution check with all required dependencies present. This outcome should be reported by object type and complexity class. A secondary outcome is regeneration success, which indicates that the agent produced an artifact but does not establish executability. The distinction is preserved from the supplied baseline, where 623 files were regenerated but only 380 regenerated outputs executed successfully.

Dependency quality should be reported as extracted dependencies, resolved internal edges, external edges, ambiguous edges, and unresolved units. For the fallback comparison, absolute counts are more informative than a single precision-like ratio because the enterprise corpus does not provide a complete gold dependency graph. When an authoritative catalog is available, future work can add precision, recall, and resolution accuracy against that catalog. Until then, provenance and explicit uncertainty are preferable to silently treating every external edge as ground truth.

Efficiency should be decomposed into baseline generation tokens, retry tokens, validation tokens, model-call count, wall-clock latency, and peak task context. Cost comparisons should be paired with correctness because a cheaper configuration that fails to create executable target objects is not an optimization. Repeated runs should report mean and standard deviation for stochastic quantities, while deterministic stages such as ANTLR decomposition and Tarjan condensation should be checked for exact output stability. Where paired tasks are used across C0-C3, paired nonparametric or bootstrap confidence intervals are preferable to unpaired aggregate comparisons because task difficulty varies strongly by object type and dependency degree.

Ablation should isolate the contribution of each architectural component. C0 measures monolithic generation. C1 introduces object decomposition without dependency context. C2 adds the dependency graph and dependency-scoped context while retaining a uniform agent. C3 adds task-conditioned dynamic agent generation and validation-driven recovery. Additional ablations can disable the LLM fallback, remove predecessor outputs from context, or force all tasks to the same model. This structure ensures that any observed gain can be attributed to a specific mechanism rather than to the cumulative effect of multiple changes introduced simultaneously.

### D. Expected Analysis by Object Type

Object type should be treated as a first-class factor because the baseline experiment already shows heterogeneous success. Tables provide a useful high-success reference class. Queries provide a low-success class for the specific specification-mediated baseline and may justify an alternative direct-translation route. Procedures and functions are important dependency-heavy classes, while indexes are useful for testing whether the graph correctly exposes schema prerequisites. Views and triggers occupy an intermediate position because their syntax may be comparatively regular while their dependencies can be non-local. Reporting one overall success percentage would hide these distinctions.

Complexity adds a second stratification dimension. The decomposition corpus contains 390 low-, 553 medium-, and 94 high-complexity units. A dynamic-agent policy should therefore be evaluated not only on the overall population but also on whether it allocates additional resources where they are useful. If high-complexity tasks consume more tokens but gain disproportionately in execution success under C3, that is evidence of effective specialization. If low-complexity tasks become more expensive without improved outcomes, the routing threshold should be adjusted.

## XVI. Discussion

The measured results support the first part of the proposed framework: heterogeneous Oracle artifacts can be transformed into a lossless, typed task set and a dependency-aware graph without making the LLM the primary structural analyzer. The strongest controlled result is not the absolute number of dependencies but the isolation of the fallback effect. Because AST dependencies remained fixed at 1,271 in both runs, the additional 496 dependencies and the elimination of 165 blind units can be attributed to the selective fallback rather than to an unrelated pipeline change.

At the same time, the results expose the next bottleneck. After fallback, 1,182 edges were still classified as external, compared with 527 resolved internal edges. Some are genuinely external objects, but others can represent unresolved internal definitions caused by schema qualification, packages, types, or objects outside the corpus. Therefore, the next graph-quality improvement should focus on resolution, not merely on extracting more names.

The dynamic-agent layer is motivated by three observations. First, the unit population is heterogeneous in complexity and dependency structure. Second, external studies show that static decomposition can incur cascading retry cost [14], while dynamic agent generation and topology selection can improve adaptability [15], [24]. Third, long-context evidence and prompt-compression research suggest that passing the entire corpus to every agent is inefficient and may reduce effective information use [3]-[5]. Dependency closures offer a principled mechanism for progressive disclosure: each agent receives only the context required by its task and predecessors.

However, this manuscript intentionally does not claim that dynamic agents already improve Oracle-to-PostgreSQL migration quality. The current empirical evidence validates decomposition, dependency recovery, cycle handling on a fixture, and the task distribution that will drive agent routing. The migration-level comparison C0–C3 remains the decisive experiment. Reporting this boundary is important because agentic orchestration papers often show strong gains in other domains, but those gains cannot be assumed to transfer unchanged to PL/SQL migration.

The newly incorporated 1,006-file experiment narrows this uncertainty without eliminating it. It demonstrates that a specification-first agent pipeline can generate a substantial number of target artifacts and that executable validation is feasible at corpus scale. More importantly, its failures align with the dependency-aware motivation: procedures, functions, and indexes were among the objects that failed most often when schema prerequisites did not match. This is evidence that schema and dependency context are material variables, but it is not yet causal evidence that the proposed graph-conditioned policy will improve them.

The baseline also reveals that task strategy may need to vary by object type. The zero-success result for queries in the supplied experiment suggests that forcing every object through the same specification-mediated generation path is not always appropriate. A mature dynamic factory should be able to route some query classes to direct dialect translation, tables to a lightweight spec-and-generate path, and complex procedural objects to a richer dependency-aware agent with executable verification. Thus dynamic generation is best interpreted as policy selection over a portfolio of migration strategies, not merely as choosing among several LLM personas.

A second implication concerns stopping rules. Generation success is only an intermediate milestone. The end-to-end funnel falls from 623 regenerated outputs to 380 executable scripts, showing that 243 of the 623 regenerated outputs (approximately 39.0%) did not pass the reported PostgreSQL execution stage. This gap justifies putting execution validation inside the orchestration loop rather than leaving it to downstream manual review. It also suggests that future experiments should report first-pass execution success, success after targeted repair, and marginal token cost per recovered task so that improvements are not hidden behind unlimited retries.

A third implication concerns provenance. The hybrid graph already records whether dependencies come from AST or LLM recovery. The same provenance discipline should extend to migration outputs: which specification version produced the target artifact, which predecessor outputs were visible, which model and tools were used, and which validations passed. This creates an audit trail suitable for both scientific analysis and practical migration review.

## XVII. Threats to Validity

Internal validity is strengthened by the controlled fallback comparison, but the corpus contains no natural cycles and therefore does not establish corpus-scale SCC frequency. Cycle condensation is currently validated on a dedicated fixture. Construct validity is limited by the distinction between external and unresolved references: an edge classified as external is not necessarily proven to be external to the complete enterprise system. The 116-file corpus is real but internal and cannot be redistributed, limiting independent replication. The LLM fallback was sequential, so its 37-minute runtime should not be interpreted as an inherent lower bound; batching, parallel calls, or caching could substantially change this cost. Finally, the dynamic-agent policy is a proposed operational mapping until the C0–C3 migration experiment is executed.

The complementary 1,006-file migration experiment has additional limitations. Its 62% regeneration and 61% post-regeneration execution rates describe the supplied implementation and dataset; they are not estimates of general Oracle-to-PostgreSQL migration accuracy. Exact per-object counts were not available for every object class, so only the explicitly reported table rate (about 85%) and query outcome (0%) are treated as quantitative class-level results. The experiment used two models and reports broad timing/token observations rather than a fully controlled model ablation. Most importantly, it did not compare the same tasks under C0–C3, so it cannot establish the causal effect of dependency-aware context or dynamic agent generation. These constraints are explicitly reflected in the interpretation of the reported results.

## XVIII. Conclusion

This paper presented a hybrid dependency-aware framework for turning Oracle SQL/PL/SQL artifacts into migration tasks and using those tasks as runtime specifications for dynamic agent generation. The deterministic-first pipeline separates syntax-derived structure from probabilistic recovery, creates a global dependency graph, collapses cycles into DAG work items, and exposes dependency-respecting phases. On 116 real Oracle files, 1,037 units were extracted with zero coverage gaps. The deterministic path produced 1,271 dependencies and 446 resolved internal edges; selective LLM fallback recovered 496 additional validated dependencies, reduced unresolved-dependency units from 165 to zero, and increased resolved internal edges to 527. A complementary specification-first migration experiment on 1,006 PL/SQL files regenerated 623 files and executed 380 of the regenerated scripts successfully in PostgreSQL 16, demonstrating both the feasibility of corpus-scale executable validation and the remaining gap between generation and operational success. The reported object-class behavior, especially the strong table result and the difficulties of queries and schema-dependent procedural objects, supports the need for differentiated routing, dependency-aware context, and task-specific validation. The proposed framework therefore includes an explicit testing, monitoring, and diagnostic layer together with a validation-driven recovery loop. The decisive next experiment remains a matched C0–C3 comparison that isolates the effect of dependency context and dynamic agent policy on executable correctness, semantic quality, token cost, latency, and retry behavior.

## Declaration on Generative AI

During the preparation of this work, the authors used ChatGPT (OpenAI) for language editing and editorial refinement. The authors reviewed and verified the resulting text and take full responsibility for the content of the publication.

## References

[1] A. Fan, B. Gokkaya, M. Harman, M. Lyubarskiy, S. Sengupta, S. Yoo, and J. M. Zhang, “Large language models for software engineering: Survey and open problems,” in Proc. 2023 IEEE/ACM Int. Conf. Software Engineering: Future of Software Engineering (ICSE-FoSE), Melbourne, Australia, 2023, pp. 31–53, doi: 10.1109/ICSE-FoSE59343.2023.00008.

[2] J. He, C. Treude, and D. Lo, “LLM-based multi-agent systems for software engineering: Literature review, vision, and the road ahead,” ACM Trans. Softw. Eng. Methodol., vol. 34, no. 5, Art. no. 124, pp. 1–30, 2025, doi: 10.1145/3712003.

[3] N. F. Liu, K. Lin, J. Hewitt, A. Paranjape, M. Bevilacqua, F. Petroni, and P. Liang, “Lost in the middle: How language models use long contexts,” Trans. Assoc. Comput. Linguistics, vol. 12, pp. 157–173, 2024, doi: 10.1162/tacl_a_00638.

[4] H. Jiang, Q. Wu, C.-Y. Lin, Y. Yang, and L. Qiu, “LLMLingua: Compressing prompts for accelerated inference of large language models,” in Proc. 2023 Conf. Empirical Methods in Natural Language Processing (EMNLP), Singapore, 2023, pp. 13358–13376, doi: 10.18653/v1/2023.emnlp-main.825.

[5] H. Jiang, Q. Wu, X. Luo, D. Li, C.-Y. Lin, Y. Yang, and L. Qiu, “LongLLMLingua: Accelerating and enhancing LLMs in long context scenarios via prompt compression,” in Proc. 62nd Annual Meeting of the Association for Computational Linguistics (ACL), Bangkok, Thailand, 2024, pp. 1658–1677, doi: 10.18653/v1/2024.acl-long.91.

[6] O. Grynets, D. Babarytskyi, and V. Lyashkevych, “Token optimization strategies for LLM-based Oracle-to-PostgreSQL migration,” arXiv preprint arXiv:2605.28557, 2026, doi: 10.48550/arXiv.2605.28557.

[7] O. Grynets, V. Lyashkevych, A. Dolichnyi, R. Piznak, T. Zelenyy, and V. Morozov, “Specification-based Code–Text–Code reengineering for LLM-mediated software evolution,” arXiv preprint arXiv:2605.25232, 2026, doi: 10.48550/arXiv.2605.25232.

[8] V. Lyashkevych, “Evolution-aware specification-driven synthesis of intelligent monitoring systems by large language models,” Informatics and Mathematical Methods in Simulation, vol. 16, no. 3, pp. 424–432, 2026, doi: 10.15276/imms.v16.no3.424.

[9] V. Lyashkevych, “Multi-drift predictive monitoring for evolving information systems,” Computer Systems and Information Technologies, no. 2, pp. 171–184, 2026, doi: 10.31891/csit-2026-2-15.

[10] M. Y. Lyashkevych, V. Y. Lyashkevych, and R. Y. Shuvar, “Security and other risks related to LLM-based software development,” Ukrainian Journal of Information Technology, vol. 7, no. 1, pp. 86–96, 2025, doi: 10.23939/ujit2025.01.086.

[11] Q. Wu et al., “AutoGen: Enabling next-gen LLM applications via multi-agent conversation,” arXiv preprint arXiv:2308.08155, 2023, doi: 10.48550/arXiv.2308.08155.

[12] O. Khattab et al., “DSPy: Compiling declarative language model calls into self-improving pipelines,” arXiv preprint arXiv:2310.03714, 2023, doi: 10.48550/arXiv.2310.03714.

[13] A. Singhvi, M. Shetty, S. Tan, C. Potts, K. Sen, M. Zaharia, and O. Khattab, “DSPy assertions: Computational constraints for self-refining language model pipelines,” arXiv preprint arXiv:2312.13382, 2023, doi: 10.48550/arXiv.2312.13382.

[14] S. Asthana, B. Zhang, C. DeLuca, H. Patel, and R. Mahindru, “Runtime-structured task decomposition for agentic coding systems,” arXiv preprint arXiv:2605.15425, 2026, doi: 10.48550/arXiv.2605.15425.

[15] Y. Wang, Z. Wu, J. Yao, and J. Su, “TDAG: A multi-agent framework based on dynamic task decomposition and agent generation,” Neural Networks, vol. 185, Art. no. 107200, 2025, doi: 10.1016/j.neunet.2025.107200.

[16] X. Liu et al., “AgentBench: Evaluating LLMs as agents,” in Proc. 12th Int. Conf. Learning Representations (ICLR), 2024, doi: 10.48550/arXiv.2308.03688.

[17] C. E. Jimenez, J. Yang, A. Wettig, S. Yao, K. Pei, O. Press, and K. R. Narasimhan, “SWE-bench: Can language models resolve real-world GitHub issues?” in Proc. 12th Int. Conf. Learning Representations (ICLR), 2024, doi: 10.48550/arXiv.2310.06770.

[18] S. Jha et al., “ITBench: Evaluating AI agents across diverse real-world IT automation tasks,” in Proc. 42nd Int. Conf. Machine Learning (ICML), PMLR, vol. 267, pp. 27134–27197, 2025, doi: 10.48550/arXiv.2502.05352.

[19] S. Pan and D. Wu, “Modular task decomposition and dynamic collaboration in multi-agent systems driven by large language models,” in Proc. IEEE Int. Symp. Parallel and Distributed Systems (ISPDS), 2025, doi: 10.1109/ISPDS67367.2025.11390999.

[20] X. Zhang et al., “Verified multi-agent orchestration: A Plan-Execute-Verify-Replan framework for complex query resolution,” arXiv preprint arXiv:2603.11445, 2026, doi: 10.48550/arXiv.2603.11445.

[21] W. Zhou, Y. Gao, X. Zhou, and G. Li, “CrackSQL: A hybrid SQL dialect translation system powered by large language models,” arXiv preprint arXiv:2504.00882, 2025, doi: 10.48550/arXiv.2504.00882.

[22] S. Kim, S. Moon, R. Tabrizi, N. Lee, M. W. Mahoney, K. Keutzer, and A. Gholami, “An LLM compiler for parallel function calling,” in Proc. 41st Int. Conf. Machine Learning (ICML), PMLR, vol. 235, pp. 24370–24391, 2024, doi: 10.48550/arXiv.2312.04511.

[23] N. Shinn, F. Cassano, E. Berman, A. Gopinath, K. Narasimhan, and S. Yao, “Reflexion: Language agents with verbal reinforcement learning,” in Advances in Neural Information Processing Systems, vol. 36, 2023, doi: 10.48550/arXiv.2303.11366.

[24] G. Yu, “AdaptOrch: Task-adaptive multi-agent orchestration in the era of LLM performance convergence,” arXiv preprint arXiv:2602.16873, 2026, doi: 10.48550/arXiv.2602.16873.

[25] L. Yue et al., “From static templates to dynamic runtime graphs: A survey of workflow optimization for LLM agents,” arXiv preprint arXiv:2603.22386, 2026, doi: 10.48550/arXiv.2603.22386.

[26] D. B. Piskala, “Spec-driven development: From code to contract in the age of AI coding assistants,” arXiv preprint arXiv:2602.00180, 2026, doi: 10.48550/arXiv.2602.00180.

[27] J. S. Ostroff, D. Makalsky, and R. F. Paige, “Agile specification-driven development,” in Extreme Programming and Agile Processes in Software Engineering, Lecture Notes in Computer Science, vol. 3092. Berlin, Germany: Springer, 2004, pp. 104–112, doi: 10.1007/978-3-540-24853-8_12.

[28] İ. E. Sarıdaş, O. Salan, A. Görçin, İ. Hökelek, and H. A. Çırpan, “RAG-driven multi-agent LLM framework with task decomposition for Beyond 5G auto-configuration,” in Proc. 32nd Int. Conf.

Telecommunications (ICT), 2026, pp. 71–76, doi: 10.1109/ICT70370.2026.11594648.

[29] F. Fournier, L. Limonad, and Y. David, “Agentic AI process observability: Discovering behavioral variability,” arXiv preprint arXiv:2505.20127, 2025, doi: 10.48550/arXiv.2505.20127.

[30] V. Lyashkevych, “Adaptive synthesis of intelligent monitoring models for LLM-modified information systems: A genetic optimisation approach,” Applied Computer Systems, vol. 31, no. 1, pp. 116–126, 2026, doi: 10.2478/acss-2026-0010.

[31] R. E. Tarjan, “Depth-first search and linear graph algorithms,” SIAM Journal on Computing, vol. 1, no. 2, pp. 146–160, 1972, doi: 10.1137/0201010.

[32] T. Parr, The Definitive ANTLR 4 Reference. Raleigh, NC, USA: Pragmatic Bookshelf, 2013. DOI: not assigned.

[33] J. Näumann, S. Keidel, A. M. Sharifloo, and M. Mezini, “Beyond BLEU: A semantic evaluation method for code translation,” arXiv preprint arXiv:2605.05282, 2026, doi: 10.48550/arXiv.2605.05282.

[34] S. Ren, D. Guo, S. Lu, L. Zhou, S. Liu, D. Tang, N. Sundaresan, M. Zhou, A. Blanco, and S. Ma, “CodeBLEU: A method for automatic evaluation of code synthesis,” arXiv preprint arXiv:2009.10297, 2020, doi: 10.48550/arXiv.2009.10297.

[35] H. Zhu, P. A. V. Hall, and J. H. R. May, “Software unit test coverage and adequacy,” ACM Computing Surveys, vol. 29, no. 4, pp. 366–427, 1997, doi: 10.1145/267580.267590.

[36] J. H. Andrews, L. C. Briand, and Y. Labiche, “Is mutation an appropriate tool for testing experiments?” in Proc. 27th Int. Conf. Software Engineering (ICSE), St. Louis, MO, USA, 2005, pp. 402–411, doi: 10.1145/1062455.1062530.